\documentclass[a4paper,twocolumn,11pt,floatfix,accepted=2024-09-24]{quantumarticle}

\pdfoutput=1

\usepackage[utf8]{inputenc}
\usepackage[english]{babel}
\usepackage[T1]{fontenc}
\usepackage{hyperref}

\usepackage{tikz}
\usepackage{lipsum}

\usepackage[numbers]{natbib}

\usepackage{amssymb}
\usepackage{xcolor}
\usepackage{natbib}
\usepackage{graphicx}
\usepackage{color}
\usepackage{braket}
\usepackage{amsmath}
\usepackage{array}
\usepackage{tabularx}
\usepackage{mathtools}
\usepackage{commath}
\usepackage{bbold}
\usepackage{svg}
\usepackage{soul}

\begin{document}
\title{Topological photon pumping in quantum optical systems}

\author{Mathias B. M. Svendsen}
\affiliation{Institut f\"ur Theoretische Physik, Universit\"at Tübingen, Auf der Morgenstelle 14, 72076 T\"ubingen, Germany}
\author{Marcel Cech}
\affiliation{Institut f\"ur Theoretische Physik, Universit\"at Tübingen, Auf der Morgenstelle 14, 72076 T\"ubingen, Germany}
\author{Max Schemmer}
\email{schemmer@lens.unifi.it}
\affiliation{Istituto Nazionale di Ottica del Consiglio Nazionale delle Ricerche (CNR-INO), 50019 Sesto Fiorentino, Italy}
\affiliation{European Laboratory for Non-Linear Spectroscopy (LENS), Università di Firenze, 50019 Sesto Fiorentino, Italy}
\author{Beatriz Olmos}
\affiliation{Institut f\"ur Theoretische Physik, Universit\"at Tübingen, Auf der Morgenstelle 14, 72076 T\"ubingen, Germany}

%\date{\today}

\begin{abstract}
We establish the concept of topological pumping in one-dimensional systems with long-range couplings and apply it to the transport of a photon in quantum optical systems. In our theoretical investigation, we introduce an extended version of the Rice-Mele model with all-to-all couplings. By analyzing its properties, we identify the general conditions for topological pumping and theoretically and numerically demonstrate topologically protected and dispersionless transport of a photon on a one-dimensional emitter chain. As concrete examples, we investigate three different popular quantum optics platforms, namely Ryd\-berg atom lattices, dense lattices of atoms excited to low-lying electronic states, and atoms coupled to waveguides, using experimentally relevant parameters. We observe that despite the long-ranged character of the dipole-dipole interactions, topological pumping facilitates the transport of a photon with a fidelity per cycle which can reach 99.9\%. Moreover, we find that the photon pumping process remains topologically protected against local disorder in the coupling parameters.
\end{abstract}

\maketitle

\section{Introduction}

The  precise control of the transport of a quantum state between different physical locations is a key ingredient for quantum information processing. Such transport can be for example implemented by moving atoms in optical tweezers~\cite{endres_atom-by-atom_2016-1,barredo_atom-by-atom_2016-1,Ordevic2021,Bluvstein2022}. However, physically displacing quantum emitters over large distances is a delicate process and, ideally, such transport should be resilient to external disturbances. In the 80's, Thouless~\cite{Thouless1983} introduced the concept of topological charge pumps as a counter-intuitive, but robust transport mechanism founded on the principles of so-called topological protection~\cite{citro_thouless_2023}. Initially, the concept of topological charge pumps or topological pumping was formulated for solid-state systems involving electron transport, but over time it has been adapted to various systems, with `charge' denoting a generic particle or excitation. In topological pumping, a center of mass displacement is introduced by adiabatically varying the system's parameters in a way that allows controlled transitions between two topologically distinct phases which are characterized by topological invariants, such as the Chern number. When realized in a physical system, these topological invariants determine many of its physical properties, which in turn show remarkable robustness against local perturbations such as disorder~\cite{thouless_quantized_1982,haldane_topological_2018}. This robustness arises from mathematical symmetries - such as inversion or time-reversal symmetry - and has already been exploited for the fabrication of more robust devices \cite{Alicea2012,Ozawa2019} and the establishment of physical standards \cite{Bachmair2003}.

\begin{figure}[]
    \centering
    \includegraphics[width=\linewidth]{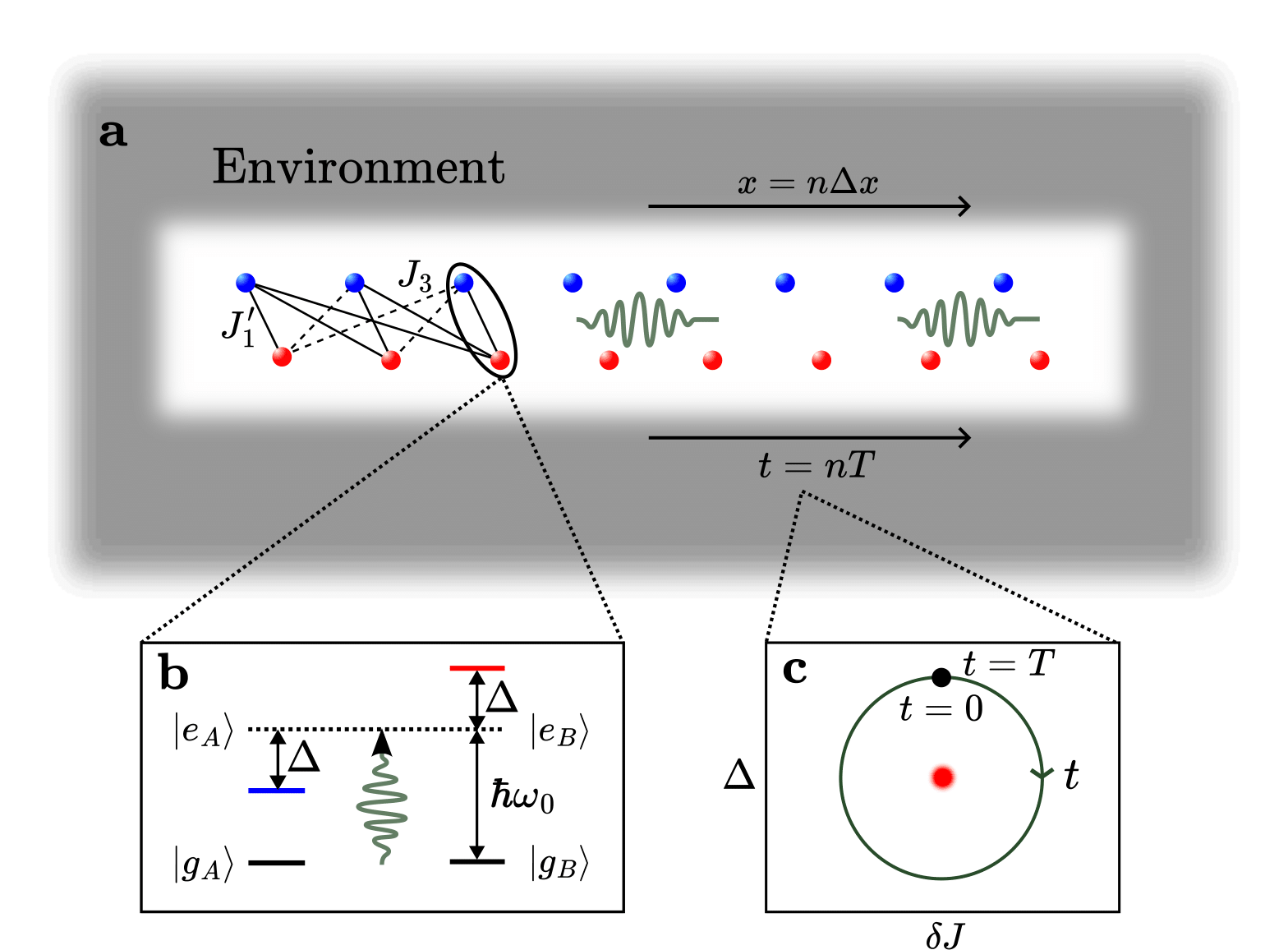}
    \caption{\textit{Topological photon pumping on a quantum optical system.} \textbf{a:} All systems we consider can be generally described as a one-dimensional chain of emitters coupled to a common (Markovian) environment. The chain is formed by $N/2$ unit cells with two sublattices, $A$ and $B$ (blue and red, respectively). \textbf{b:} The emitters are considered to be identical two-level systems with ground ($\ket{g}$) and excited ($\ket{e}$) states separated by an energy $\hbar \omega_0$. \textbf{c:} By varying time-dependently the parameters of the system (here generically called $\Delta$ and $\delta J$) in a periodic manner around a topological degeneracy point (red dot), a single photon is transported (topologically pumped) across the lattice minimizing its dispersion.}
    \label{Fig:Fig0}
\end{figure}

Only over the last decade, engineered systems such as waveguide arrays \cite{Kraus2012,Verbin2015,Ke2016,Cerjan2020}, electronic systems \cite{grinberg_robust_2020,xia_experimental_2021} and cold atoms systems \cite{lohse_thouless_2016,nakajima_topological_2016,Lu2016} offered sufficient dynamical control of the system's parameters to experimentally realize topological charge pumps. The displacement arises when the parameters are varied adiabatically around a topological degeneracy point. When an energy band is fully or uniformly occupied, the average displacement is given by an integer number of lattice sites~\cite{lohse_thouless_2016,nakajima_topological_2016}. For inhomogeneous occupation of the bands, the displacement per cycle can be adjusted~\cite{Lu2016,ke_topological_2020}. In all these systems, the Hamiltonians are either non-interacting or display short range couplings (nearest neighbor hopping on a lattice, for example) for which the topological characterization and classification is relatively well understood \cite{Kitaev2009,Ryu2010,Chiu2016,Chaudhary2021}. Recent advancements have extended this understanding to more complex non-Hermitian Hamiltonians \cite{Zeuner2015,Dangel2018,Lieu2020,Bergholtz2021} and Hamiltonians with on-site interactions \cite{viebahn2023interactioninduced, Zhu2024, walter_quantization_2023}. Conversely, very little is known about the fate of topological properties in systems with long-range couplings. Recent works indicate that the effect of such long-ranged character varies from system to system \cite{Pocock2019,Perez2019,Appugliese2022,Dias2022}. Strikingly, it has been recently shown that some of the topological properties survive even in the presence of infinite range couplings, which can be found in, e.g. cavity and waveguide QED setups \cite{Bello2019,Mcdonnell2022,Allard2023}.

Understanding the interplay between topology and long-range couplings extends well beyond the fundamental interest in condensed matter physics. This affects particularly the field of quantum optics, where systems inherently exhibit long-range dipole-dipole interactions. These systems constitute nowadays the vanguard of platforms for quantum simulation and quantum information processing with quantum emitters \cite{Gross2017,chang_colloquium_2018}, which can be realized with cold atoms \cite{Deleseleuc2019,ordevic_entanglement_2021,bluvstein_logical_2024}, molecules \cite{Blackmore2019,bao_dipolar_2023,holland_-demand_2023} or resonators \cite{Hafezi2014,Ozawa2019,Adiyatullin2023} placed in different one- two- and three-dimensional geometries. The coupling between these few-level emitters to a common environment can occur naturally, like in the case of the free space radiation field \cite{Bettles2017,Perczel2017,Cech2023,dePaz2023,gonzalez-tudela_lightmatter_2024} or be engineered via optical and microwave cavities \cite{colombe_strong_2007,kuhn_optical_2002}, tapered optical waveguides \cite{vetsch_optical_2010,nayak_nanofiber_2018}, and photonic crystals \cite{goban_atomlight_2014,Proctor2020,ordevic_entanglement_2021,bouscal_systematic_2024}, among others (see Fig.~\ref{Fig:Fig0}a). It is therefore an open question whether the promises of topology and in particular topological pumping can be applied in such systems.

In this work, we investigate and obtain the conditions necessary to realize topological pumping in systems with long-range couplings (Sec.~\ref{sec:extendedRiceMele}). Moreover, we apply these results to theoretically demonstrate for the first time the topological pumping of a photon embedded in a one dimensional chain of emitters, i.e., how its position can be moved in a controlled way, while also being topologically protected against local disorder. Crucially, we take care that this transport is also dispersionless, such that not only the center of the photon wave packet is transported, but also its shape is conserved over several pump cycles, allowing for a high fidelity reaching 99.9\% per cycle. We study this transport mechanism on three explicit platforms (Sec.~\ref{sec:Top_pumping_in_Qotpics}), namely Ryd\-berg atoms, atoms excited to low-lying electronic states, and atoms coupled to nanophotonic waveguides, which represent a relevant portion of the currently implemented experiments within the quantum optics community. Finally, we show that the topological transport mechanism is robust against dissipation and disorder (Sec.~\ref{sec:robustness}).

\section{Topological pumping in one dimension}

Let us start by reviewing the mechanism behind topological pumping. We consider a one-dimensional system governed by a generic Hamiltonian $\hat{H}$ that may host topological phases. The parameters of the Hamiltonian can be varied in time, and we consider a cyclic variation such that at the end of the cycle period $T$ the Hamiltonian is again the initial one, i.e., $\hat{H}(t=T)=\hat{H}(t=0)$. Furthermore, we consider an adiabatic evolution, such that after each cycle the initial state $\ket{\Psi_0}=\sum_m f_m \ket{u_m(t=0)}$, which is a superposition of the Hamiltonian's eigenstates $\ket{u_m(t=0)}$ with amplitude $f_m$, reaches a final state that reads $\ket{\Psi_T}=\sum_m f_m e^{i\gamma_m}\ket{u_m(t=0)}$. The geometric phase $\gamma_m$ picked up by every eigenstate at the end of the cycle is the so-called Berry phase, a quantity reflecting the geometrical properties of the system~\cite{Berry1984,Cohen2019} defined as
\begin{equation}
    \gamma_m=\mathrm{i}\int_0^T\mathrm{d}t\langle u_{m}(t)|\partial_t u_m(t)\rangle,\label{eq:Berry_m}
\end{equation}
where $\ket{u_m(t)}$ are the instantaneous eigenstates of the Hamiltonian. This geometric phase can cause a displacement of the average position of the state in real space after the cycle given by $\Delta x=\langle \hat{x}\rangle_T-\langle\hat{x}\rangle_0$, where $\langle\hat{x}\rangle_t\equiv \bra{\Psi_t}\hat{x}\ket{\Psi_t}$, which constitutes the so-called charge pumping.

In one-dimensional periodic systems like the ones we will consider in this paper, it is convenient to work in quasi-momentum space. The eigenstates of the Hamiltonian are Bloch waves of the form $\ket{\psi_{nk}(x,t)}\propto e^{\mathrm{i}kx}\ket{x}\otimes\ket{u_{nk}(t)}$, parametrized by their band index $n$ and the crystal quasi-momentum $\hbar k$. Here, $\ket{x}$ is the position eigenstate ($\hat{x}\ket{x}=x\ket{x}$), and $\ket{u_{nk}(t)}$ the cell-periodic part of the wave function. We consider the state of the dynamics at any time to be a superposition of these Bloch waves, i.e., $\ket{\Psi_t(x)} \propto \sum_k f(k) \ket{\psi_{nk}(x,t)}$, where $f(k)$ is the quasi-momentum distribution amplitude. As detailed in Appendix~\ref{app:displacement}, one finds, for a general $f(k)$ distribution, that the displacement of the average position after a cycle reads
\begin{equation}
    \Delta x   =  \frac{a}{2\pi} \int_0^T \mathrm{d}t \int_\mathrm{FBZ}  |f(k)|^2\Omega^n_{tk} \mathrm{d}k\label{eq_generalized_Berry_curv},
\end{equation}
with $a$ being the period of the crystal, FBZ denoting the First Brillouin Zone and where 
\begin{equation}
    \Omega^n_{tk} = -2\text{Im}\{\langle\partial_t u_{nk}(t)|\partial_k u_{nk}(t)\rangle\}
\end{equation}
is the so-called Berry curvature of the $n$-th band \cite{King-Smith1993,Cohen2019}. With a few notable exceptions \cite{Lu2016,Hayward2018,Unanyan2023}, typically the literature of topological pumping considers a filled band or a single-particle wave packet that is delocalized in momentum space, such that $f(k)$ is a constant function. In this case, the charge displacement is quantized in the form of $\Delta x = a C_n$, where $C_n$ is an integer called the Chern number of the band, defined as the integral of the Berry curvature over the two-dimensional torus formed by the FBZ and the periodic cycle in time. Let us remark that for the displacement to be nonzero, a topological degeneracy point, i.e. the point in parameter space where the topological invariant changes, must be placed inside this torus \cite{xiao_berry_2010}. 
Finally, note that, in general, also a dynamical phase may be present, which is proportional to the usual group velocity. This in turn may lead to a further displacement of the average position, given by the convolution of the group velocity with the quasi-momentum distribution, integrated over the FBZ. In this paper, however, we are interested in the \textit{anomalous displacement} arising from the topological properties and an adiabatic change of the system. We keep this effect dominant by
restricting ourselves to symmetric wave packets placed in the center of the FBZ of a time-reversal symmetric Hamiltonian, which leads to a zero contribution to the displacement of the dynamical phase.

%\textcolor{red}{Finally, note that, in general, also a dynamical phase may be present, which is proportional to the group velocity. This in turn may lead to a further displacement of the average position, given by the convolution of the group velocity with the quasi-momentum distribution, integrated over the FBZ. In this paper, however, we restrict ourselves to symmetric wave packets placed in the center of the FBZ of a time-reversal symmetric Hamiltonian,which lead to a zero contribution to the displacement of this dynamical phase}.

\begin{figure}[h]
    \centering
    \includegraphics[width=\linewidth]{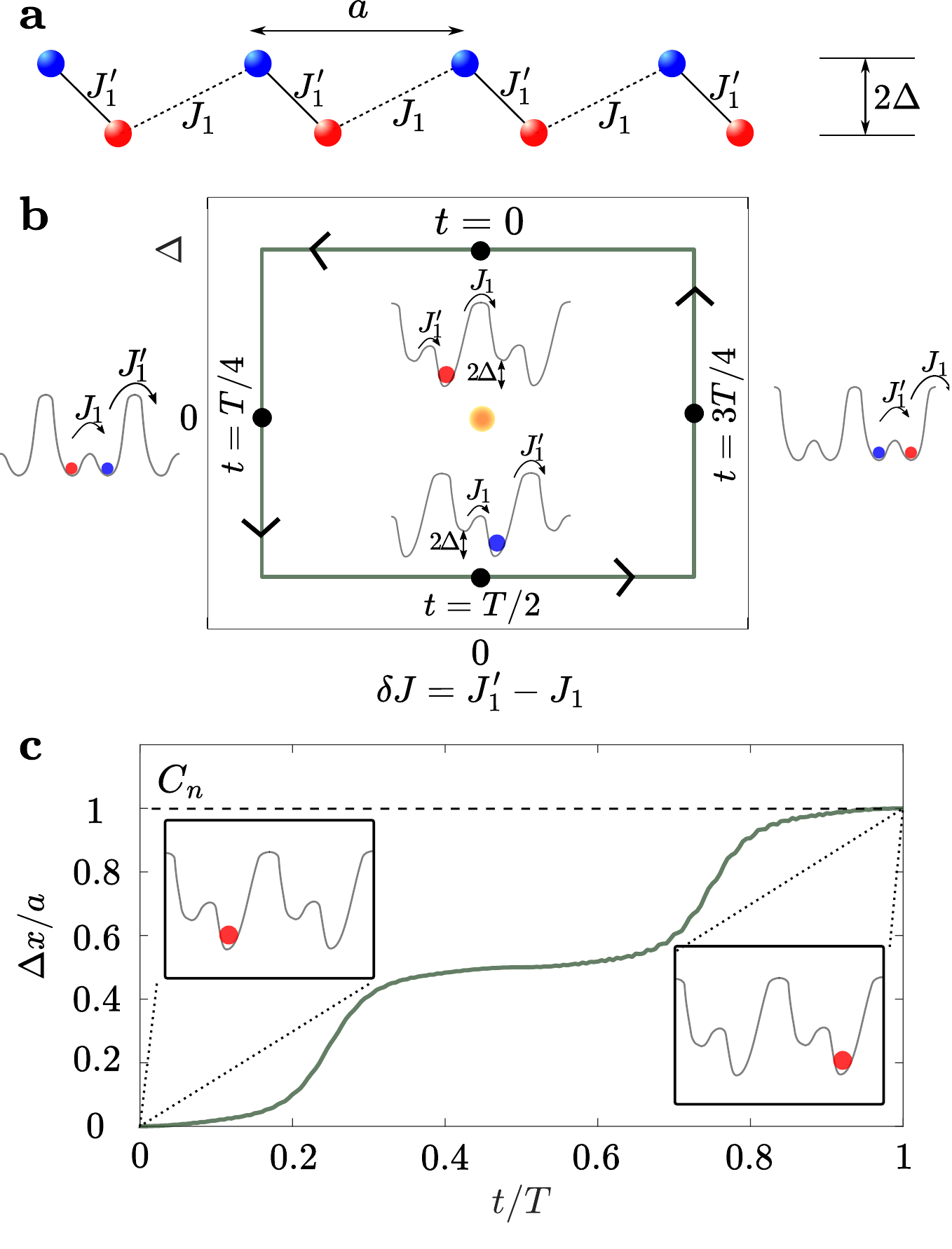}
    \caption{\textit{Topological pumping in the Rice-Mele model.} \textbf{a:} Sketch of the Rice-Mele model. $N/2$ cells formed by two sublattices, $A$ and $B$ (blue and red, respectively) are placed periodically with nearest neighbor distance $a$. Fermions can hop between sublattices (with hopping parameters $J'_1$ and $J_1$), and the sublattices are energetically offset by $2\hbar\Delta$. \textbf{b:} Adiabatically varying the parameters $\delta J=J_1'-J_1$ and $\Delta$ in time in a closed cycle around the topological singularity point [orange dot at $(\Delta,\delta J)=(0,0)$], one can achieve topologically protected displacement of the fermions by an amount $\Delta x=a$ per cycle, as shown in panel \textbf{c}.}
    \label{TopPumpExp}
\end{figure}

\subsection{Rice-Mele model}

The emblematic model Hamiltonian for charge pumping in one dimension is the so-called Rice-Mele model. Here, a one dimensional chain with $N$ sites is formed by two sublattices, $A$ and $B$. Fermions can hop between sublattices in the same cell and between neighboring cells with hopping parameters $J'_1$ and $J_1$, respectively (see Fig.~\ref{TopPumpExp}a). In addition, an energy offset $2\hbar\Delta$ is introduced between the two sublattices. This system is described by the Hamiltonian
\begin{equation}
\begin{split}
    \hat{H} =& \hbar\sum_{q=1}^{N/2}(J_1' \hat{a}_q^\dagger \hat{b}_{q}+J_1 \hat{b}_q^\dagger \hat{a}_{q+1})+\text{h.c.} \\ &+ \hbar\Delta \sum_{q=1}^{N/2} (\hat{a}^\dagger_q \hat{a}_q-\hat{b}^\dagger_q \hat{b}_q),
\end{split}
\label{RM}
\end{equation}  
where $\hat{a}_q$ and $\hat{b}_q$ ($\hat{a}_q^\dagger$ and $\hat{b}_q^\dagger$) are fermionic annihilation (creation) operators on sublattice $A$ and $B$ respectively. When the energy offset $\Delta=0$, this Hamiltonian represents the celebrated Su–Schrieffer–Heeger (SSH) model. The SSH model hosts topological phases associated with a topological invariant -- the winding number $\nu$. For $|J_1'|>|J_1|$ the system is in a topologically trivial phase with winding number $\nu = 0$, while for $|J_1'|<|J_1|$ a topologically non-trivial phase with winding number $\nu = 1$ occurs.

In the Rice-Mele model where both $J_1$ and $J_1'$ have the same sign, charge (fermion) pumping can be implemented when the parameters are varied in a cycle around the topological degeneracy point at $(\Delta, \delta J=J_1'-J_1)=(0,0)$ as sketched in Fig.~\ref{TopPumpExp}b. Such a pumping scheme is robust against perturbations which deform the path, as long as it encircles the singularity. As sketched in Fig.~\ref{TopPumpExp}c, when $f(k)$ constant in $k$, the center of mass of the wave packet is displaced after one cycle exactly by $\Delta x=a$, with $a$ being the distance between two neighboring unit cells. This quantized displacement is consistent with $\Delta x=a C_n$ since the Chern number of the lowest band is indeed  $C_n=+1$~\cite{xiao_berry_2010}.

\subsection{Topological pumping and dispersion}\label{sec_classificaiton_topological_pumping}

While Equation \eqref{eq_generalized_Berry_curv} tells us about the displacement of the center of mass of a given initial state, it does not contain any other information about its evolution in real space. However, in this paper we are not only interested in realizing topological charge pumping of the center of mass, but we aim to realize this in a dispersionless way, i.e. keeping the value of the fidelity $\mathcal{F}(T) =|\langle\Psi_0(x+\Delta x)|\Psi_T(x)\rangle|^2$ as close to one as possible. Hence, here we analyze this dispersion for different combinations of initial states and Berry curvatures.

Let us start by analysing the dispersion of the initial wave packet in real space $\ket{\Psi_0(x)} \propto \sum_k f(k) \ket{\psi_{nk}(x,0)}$. After a full period, the state becomes $\ket{\Psi_T(x)} \propto \sum_k f(k) e^{\mathrm{i}\gamma(k)}\ket{\psi_{nk}(x,0)}$, where 
\begin{equation}
    \gamma(k)=\mathrm{i}\int_0^T\mathrm{d}t\langle u_{nk}(t)|\partial_t u_{nk}(t)\rangle,
\end{equation}
as introduced in \eqref{eq:Berry_m}, is the geometric Berry phase for eigenstates in momentum space. Importantly, note that if this phase is linear in $k$, i.e. $\gamma(k)\approx k \Delta x$, for all $k$ where $f(k)\neq 0$, then $\ket{\Psi_T(x)} \approx \ket{\Psi_0(x+\Delta x)}$, so that the wave function remains unchanged except for a shift of its center. Otherwise, the wave packet disperses and the fidelity quickly drops below one. Since the time-integrated Berry curvature is directly given by the gradient of the Berry phase
\begin{equation}
    {\cal W}_k^n\equiv \int_0^T \mathrm{d}t \,\Omega^n_{tk}=-\partial_k \gamma(k),
\end{equation}
the dispersion of the wave packet is minimal when the integrated Berry curvature is constant, or `flat' in quasi-momentum space. 

\begin{figure}[t]
    \centering
    \includegraphics[width=\linewidth]{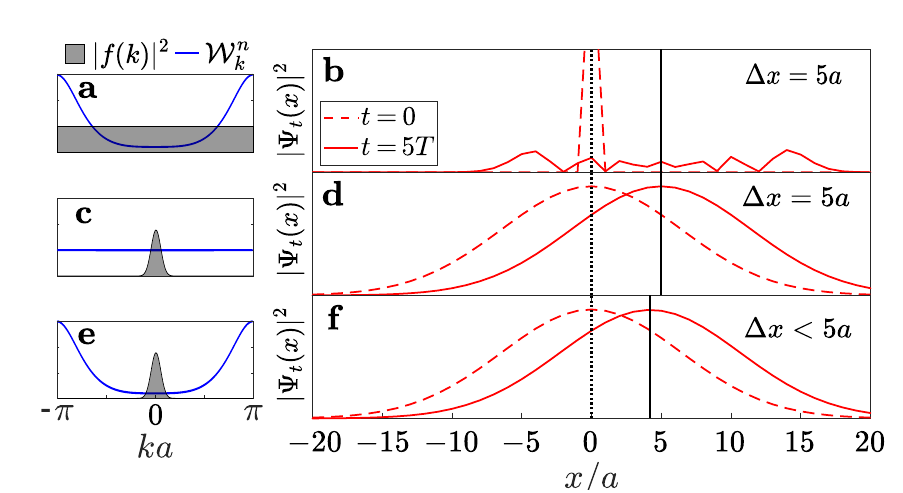}
    \caption{\textit{Topological pumping and dispersion.} \textbf{Left column:} Time-integrated Berry curvature $\mathcal{W}^n_k$ and quasi-momentum distribution density $|f(k)|^2$. \textbf{Right column:} Spatial probability density of the state, $|\Psi_t(x)|^2 \equiv {}\langle \Psi_t(x)| \Psi_t(x)\rangle $, at times $t=0$ and $t=5T$, obtained from the numerical simulation of the Schrödinger equation under a time-dependent Hamiltonian of the form \eqref{RM}, with the quasi-momentum distribution density and time-integrated Berry curvature shown in the corresponding left panel. The vertical points and solid lines indicate the center of mass of the wave packet at $t=0$ and $5T$, respectively. \textbf{a} and \textbf{b:} In standard topological pumping, where the band is uniformly filled, the non-flat shape of ${\cal W}_k^n$ across the FBZ leads to dispersion of the wave packet in real space. Dispersionless topological pumping can be achieved either by making ${\cal W}_k^n$ constant (\textbf{c} and \textbf{d}), or by constraining $|f(k)|^2$ to a domain in momentum space where ${\cal W}_k^n$ is approximately constant (\textbf{e} and \textbf{f}).}
    \label{TopPumpTypes}
\end{figure}

When the initial state is such that $\left|f(k)\right|^2$ constant for all $k$ (Fig. \ref{TopPumpTypes}a)~\cite{citro_thouless_2023,Unanyan2023}, the charge displacement is quantized in the form of $\Delta x = a C_n$. However, in general ${\cal W}_k^n$ is not flat, which leads to a quick dispersion of an initially localized wave packet in real space (see Fig. \ref{TopPumpTypes}b). Conversely, as discussed above, if ${\cal W}_{k}^n$ is constant throughout the full FBZ (see Fig. \ref{TopPumpTypes}c), quantized and dispersionless pumping will be achieved independently of the specific shape of the wave packet in $k$ space (see Fig. \ref{TopPumpTypes}d). This is, however, hard to achieve in practice in a real physical system, and even harder for a long-ranged Hamiltonian, which develops sharp structures in the Berry curvature.

In this paper, we resort to an intermediate approach, typically called \textit{geometric} pumping \cite{Lu2016}. For each system, we identify a path that leads to a time-integrated Berry curvature ${\cal W}_k^n$ that is constant \textit{on a finite region of the FBZ}. We consider then a localized quasi-momentum distribution $f(k)$ such that $f(k)\neq 0$ only in the region where ${\cal W}_k^n$ is constant (see Fig. \ref{TopPumpTypes}e)~\cite{ke_topological_2020}. This condition is much easier to fulfill than a fully flat ${\cal W}_k^n$ \cite{Unanyan2023}, and allows an approximately dispersionless transport of the wave packet in real space. Instead of obtaining a displacement $\Delta x$ that is quantized, its value will depend on the specific geometrical properties of the underlying Hamiltonian (e.g., short versus long-ranged) as well as on the choice of the pumping cycle, since this determines the shape and value of ${\cal W}_k^n$ (see Fig. \ref{TopPumpTypes}f). By choosing different protocols one can then achieve dispersionless transport of the wave packet where $\Delta x$ is larger or smaller than the one achieved in topological pumping.

%We refer the reader to Appendix \ref{app:adiabaticity} for further details \textbf{Now we removed this appendix}.
%Throughout the paper, we fulfil both conditions with $T\Delta E = 300$.

\section{Long-range hopping in quantum optics: the extended Rice-Mele model}\label{sec:extendedRiceMele}

In this paper we focus on the realization of topological pumping using quantum optics platforms and, in this Section, we present the general equation that describes the dynamics of these systems. We then introduce a mapping onto a generalized version of the Rice-Mele model with long-range hopping parameters, which we refer to as the \textit{extended Rice-Mele model}, and discuss its topological properties.

All three quantum optics platforms we discuss here share the following description (see Fig.~\ref{Fig:Fig0}a): An ensemble of $N$ two-level emitters in the Lamb-Dicke regime with ground state $\left|g\right>$ and excited state $\left|e\right>$ separated by an energy $\hbar\omega_0$, are coupled to a common environment\footnote{Here, the zero-temperature radiation field, either in free-space or with imposed boundary conditions that arise from the presence of a nanophotonic structure, such as a nanofiber.}.
Within the Born-Markov and secular approximations, the quantum master equation that describes the evolution of the emitter's internal degrees of freedom is
\begin{equation}
    \dot{\hat{\rho}}=-\frac{\mathrm{i}}{\hbar}\left[\hat{H},\hat{\rho}\right]+{\cal D}(\hat{\rho}),
    \label{eq:MEQ}
\end{equation}
where $\hat{\rho}$ is the density operator. The first term of this master equation represents the coherent evolution of the system under the Hamiltonian $\hat{H} = \hat{H}_\mathrm{dd}+\hat{H}_\Delta$ which reads
\begin{equation}
    \hat{H}=\hbar\!\sum_{i\neq j=1}^N V_{ij}\hat{\sigma}_i^\dag\hat{\sigma}_j+\hbar\Delta\sum_{j=1}^N (-1)^j\hat{\sigma}_j^\dag\hat{\sigma}_j, \label{eq:Hamiltonian_QO}
\end{equation}
with the spin-$1/2$ ladder operators $\hat{\sigma}_i=\left|g_i\right>\left<e_i\right|$ and $\hat{\sigma}_i=\left|e_i\right>\left<g_i\right|$. This Hamiltonian drives an exchange (dipole-dipole) interaction between the emitters. The strength and range of the exchange, encoded in the matrix elements $V_{ij}$, will be determined by the system at hand. The second term, $\hat{H}_\Delta$, is an on-site potential which leads to an offset $2\hbar\Delta$ between even and odd sites (see Fig.~\ref{Fig:Fig0}b). The dissipation in the system (second term of the master equation) also depends on the system at hand, and its action on the dynamics will be considered separately for each system in Sec.~\ref{sec:dissipation}. 

We assume that the emitters are placed on a bipartite one-dimensional lattice (see Fig.~\ref{Fig:Fig0}a). A Jordan-Wigner transformation of the form
\begin{eqnarray}
    \hat{\sigma}_j&=&e^{\mathrm{i}\pi\sum_{k=1}^{j-1}\hat{f}_k^\dag \hat{f}_k}\hat{f}_j,\\
    \hat{\sigma}_i^\dag&=&e^{-\mathrm{i}\pi\sum_{k=1}^{i-1}\hat{f}_k^\dag \hat{f}_k}\hat{f}_i^\dag,
\end{eqnarray}
allows us to express the spin operators in terms of the spinless fermionic creation and annihilation operators $\hat{f}_k^\dag$ and $\hat{f}_k$, respectively. Applying this transformation to Hamiltonian~\eqref{eq:Hamiltonian_QO}, and imposing the constraint of staying in the single excitation (fermion) sector, it becomes
\begin{equation}
    \hat{H}=\hbar\sum_{i\neq j=1}^N V_{ij}\hat{f}_i^\dag \hat{f}_j+\hbar\Delta\sum_{j=1}^N (-1)^j\hat{f}_j^\dag \hat{f}_j\label{eq_H_f_ij}.
\end{equation}
In the next step, we make the bipartite lattice explicit by denoting $\hat{a}_q=\hat{f}_{2q-1}$ and $\hat{b}_q=\hat{f}_{2q}$ where $q=1\dots N/2$ labels the unit cells. Accordingly, we categorize $V_{ij}$ into three different hopping
parameters: from sublattice $A$ to $B$, from $B$ to $A$, and within the same sublattice, which we denote $J'_{2p-1}$, $J_{2p-1}$ and $J_{2p}$, respectively, where $p$ denotes again the unit cell (see Fig. \ref{fig:sketches}a). The extended Rice-Mele Hamiltonian is then given by
\begin{equation}
\begin{split}
 &\hat{H} = \hbar\bigg\{\sum_{q=1}^{N/2}\sum_{p=1}^{N/2-q}\bigg[J'_{2p-1}\hat{a}_q^\dagger \hat{b}_{q+p-1}\\\label{ExtendedRM} &+J_{2p-1}\hat{b}_q^\dagger \hat{a}_{q+p}+J_{2p}(\hat{a}_q^\dagger \hat{a}_{p+q}+\hat{b}_q^\dagger \hat{b}_{p+q})\bigg]\\&+ \!J'_{N-1}\hat{a}_1^\dagger \hat{b}_{N/2}\!+\!\text{h.c.}\!\bigg\} \!+\! \hbar\Delta \!\sum_{j=1}^{N/2} (\hat{a}^\dagger_j\hat{a}_j\!-\!\hat{b}^\dagger_j\hat{b}_j).
\end{split}
\end{equation}  
%Note, that in the limit of hopping only between nearest neighbors, this Hamiltonian reduces to the standard Rice-Mele model \eqref{RM}.
Note that, when restricting hopping to solely nearest neighbors, this Hamiltonian reduces to the standard Rice-Mele mode~\eqref{RM}.

\subsection{Topological properties of the extended Rice-Mele model}

Let us here analyze the bulk properties of the extended Rice-Mele model. Considering periodic boundary conditions, Hamiltonian \eqref{ExtendedRM} may be written in terms of its irreducible matrix form in momentum space $h(k)$ as
\begin{equation}
    \hat{H} = \sum_{\substack{k\in FBZ\\ \alpha,\beta\in A,B}} \hat{c}^\dag_{\alpha}(k) h_{\alpha\beta}(k) \hat{c}_{\beta}(k), 
    \label{BlochHamiltonian}
\end{equation}
where 
\begin{equation}
    \hat{c}_\alpha(k)=\frac{1}{\sqrt{\lfloor N/4\rfloor}}\sum_{p=1}^{\lfloor N/4\rfloor} e^{-\mathrm{i}kpa}\,\hat{c}_\alpha(p),
\end{equation}
with $\hat{c}_A(p)\equiv \hat{a}_p$ and $\hat{c}_B(p)\equiv \hat{b}_p$ and
\begin{equation}
    h(k) = 
    \begin{pmatrix}
        n_0(k)+\hbar\Delta && n(k) \\ 
        n^*(k) && n_0(k)-\hbar\Delta
    \end{pmatrix}
    .
\end{equation}
Here, we have defined the functions
\begin{equation}
\begin{split}
    &n_0(k) = 2\hbar\sum_{p=1}^{\lfloor N/4 \rfloor} J_{2p}\cos(k p a), \\
    &n(k) = \hbar\!\!\sum_{p=1}^{\lfloor N/4 \rfloor}\!\!\!\left(J_{2p-1}e^{i k p a}\!+\!J'_{2p-1}e^{-i k (p-1) a}\right)
\end{split}
\end{equation}
that contain the intra-sublattice and inter-sublattice hopping parameters respectively. Diagonalizing $h(k)$ we obtain the dispersion relation $E_{\pm}(k) = n_0(k)\pm \sqrt{|n(k)|^2+\Delta^2}$ for the upper ($+$) and lower ($-$) band, respectively. The extended Rice-Mele Hamiltonian with $\Delta\neq 0$ does not host topological phases. However, for vanishing energy offset $\Delta = 0$, the Hamiltonian reduces to the extended SSH Hamiltonian~\cite{Perez2019,Mcdonnell2022}. This Hamiltonian can be allocated to the BDI symmetry class, which possesses a $\mathbb{Z}$-type topological invariant in one dimension, like the standard SSH model \cite{Ryu2010}, when the sublattice symmetry is conserved, i.e. when $n_0(k)=0$, which can be accomplished by setting $J_{2p}=0$ for all $p$. Note, however, that even though we impose here, where possible, sublattice symmetry, the model described by \eqref{ExtendedRM} is inversion symmetric and hence $J_{2p}=0$ is not a necessary condition for the pumping to take place.

To calculate the topological degeneracy points in the extended Rice-Mele model we look for the points in the FBZ where the energy gap closes and the argument of the complex function $n(k)$ is discontinuous. The imaginary part of $n(k)$ is equal to zero when $k=\pm \pi/a$ and $k=0$. At those points, also the real part of $n(k)$ is zero when
\begin{equation}
    \sum_{p=1}^{\lfloor N/4 \rfloor}(-1)^{p+1}(J_{2p-1}'-J_{2p-1})=0
\end{equation}
for $k=\pm\pi/a$ and
\begin{equation}
    \sum_{p=1}^{\lfloor N/4 \rfloor}(J_{2p-1}'+J_{2p-1}) = 0
\end{equation}
for $k=0$, making the argument of $n(k)$ undefined and the energy gap zero. Note, that there are further specific fine-tuned combinations of the parameters $J_{2p-1}$ and $J'_{2p-1}$ that lead to topological degeneracy points at other points of the FBZ. However, in the pumping cycles and physical systems that we propose in this paper the gap only closes at $k=\pm\pi/a$. Thus, we introduce the extended hopping parameters
\begin{equation}
    \begin{split}
       \bar{J}' &= \sum_{p=1}^{\lfloor N/4 \rfloor}(-1)^{p+1}J_{2p-1}', \\
       \bar{J} &= \sum_{p=1}^{\lfloor N/4 \rfloor}(-1)^{p+1}J_{2p-1},
    \end{split}
\end{equation}
and realize cyclic variations in the extended Rice-Mele Hamiltonian parameters such that the curve $\xi(t) = ( \Delta,\delta\bar{J}=\bar{J}'-\bar{J})$ encircles the origin.

\section{Topological photon pumping in quantum optical systems}\label{sec:Top_pumping_in_Qotpics}

\begin{figure}[h!]
    \centering
    \includegraphics[width=\linewidth]{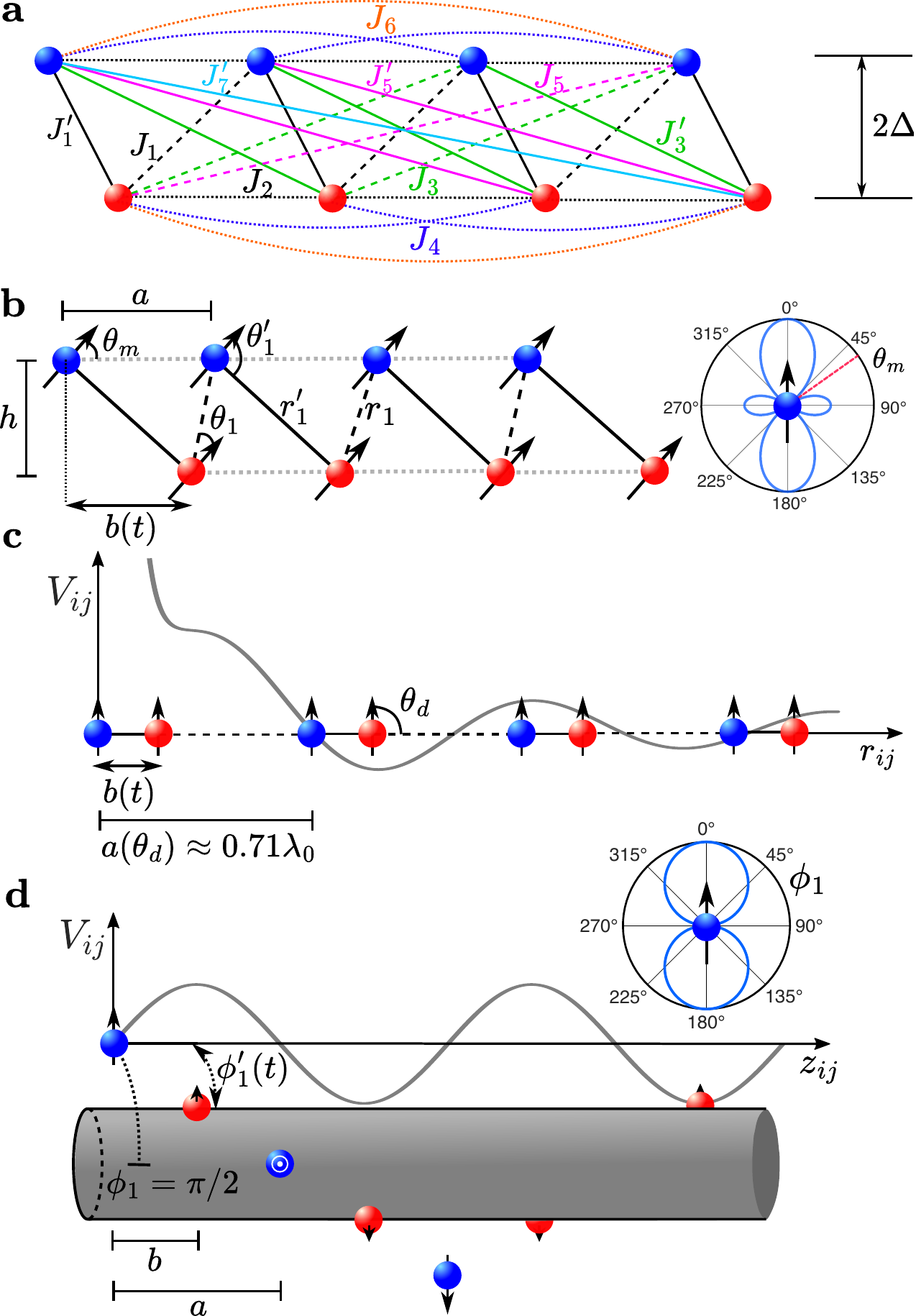}
    \caption{\textit{Realizations of the extended Rice-Mele model.} \textbf{a:} In the extended Rice-Mele model, we consider in general non-zero hopping parameters between all sites. Here, an example for a system with $N=8$ sites. \textbf{b:} In the Ryd\-berg system, the two sublattices are separated by a perpendicular distance $h$ and a horizontal offset $b(t)$, which can be changed in time to realize the photon pumping. Here, the nearest neighbor hoppings dominate and we exploit the dependence of the couplings on dipole moment orientation (right panel) to make the intra-sublattice hoppings ($J_{2p}$) negligible. \textbf{c:} Similarly, on a dense chain of atoms excited to low-lying electronic states, $J_2$ is put to zero by fixing a value of $a$ that depends on the angle $\theta_\mathrm{d}$ of the dipoles, while the rest of the hoppings (determined by $V_{ij}$) are non-zero in general. \textbf{d:} In a lattice of atoms coupled to a waveguide (here a cylindrical nanofiber), the atoms of sublattice $A$ are fixed forming a helix around the fiber, while the atoms of sublattice $B$ change their relative angle to sublattice $A$, $\phi'_1(t)$, time-dependently to realize the pumping. Sublattice symmetry is preserved by choosing $a=\pi/\beta$, where $\beta$ is the wave vector of the light propagating inside the fiber.}
    \label{fig:sketches}
\end{figure}

Having set up the stage  for the realization of dispersionless topological pumping  in the extended Rice-Mele model (see Fig.~\ref{fig:sketches}a), we now examine three quantum optics platforms of current experimental relevance. These platforms enable the pumping of photons stored in the excited state of two-level emitters and encompass: a Ryd\-berg lattice, a dense lattice gas of atoms in low-lying electronic states, and atoms coupled to a nanophotonic waveguide as sketeched in Fig.~\ref{fig:sketches}b - d.
These three platforms all present long-range interactions, but the different shape and strength of the couplings lead to differences in their effective range. E.g. in a Ryd\-berg lattice system the nearest neighbor hopping parameters dominate the dynamics, while in the waveguide system all atoms are coupled to all other atoms with similar strength, such that excitation hopping stretches across all distances.

In all cases, we numerically simulate the dynamics of the Schrödinger equation with a time-dependent Hamiltonian of the form \eqref{ExtendedRM} for a system with open boundary conditions.
%\textcolor{red}{In all cases, we perform a numerical simulation of the dynamics given by the solution of the Schrödinger equation with a time-dependent Hamiltonian of the form \eqref{ExtendedRM} for a system with open boundary conditions.
Note that the cycle period $T$ needs to be sufficiently long to ensure adiabaticity. This is fulfilled when $T\gg \hbar/\Delta E$, where $\Delta E$ is the minimum energy gap between neighboring bands during the cycle.
%Note that the cycle period $T$ needs to be slow enough to ensure adiabaticity. This is fulfilled when $T\gg \hbar/\Delta E$, where $\Delta E$ is the minimum energy gap between neighboring bands during the cycle.
Conversely, our transport protocols aim to minimize dispersion, requiring that the cycle is fast enough to prevent non-linearities in the Berry phase $\gamma(k)$ from introducing wave packet dispersion
%On the other hand, for the transport to be as dispersionless as possible we need to consider an additional constraint: The cycle needs to be fast enough such that the small non-linearities inevitably present in the Berry phase $\gamma(k)$ do not lead to the dispersion of the wave packet.
Throughout the paper, we have found cycle periods that approximately optimize these two conditions (we fix $T=300/\Delta E$), although this still leads to a small degree of dispersion.

\subsection{Short-range hopping: Ryd\-berg atoms}

\begin{figure*}[t]
    \centering
    \includegraphics[width=\linewidth]{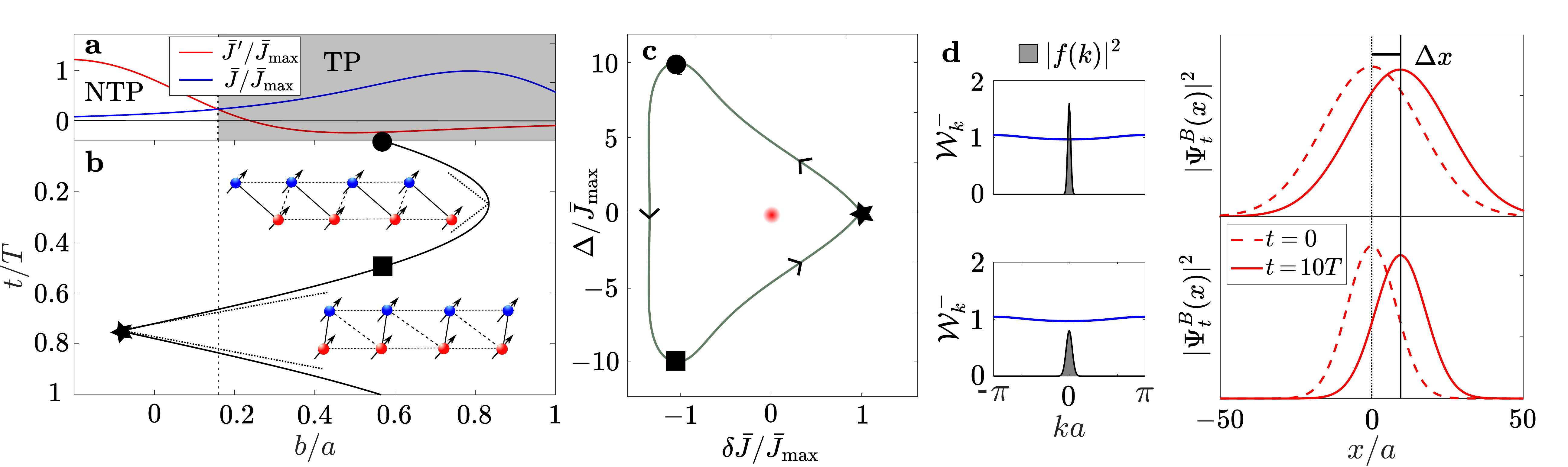}
    \caption{\textit{Photon pumping on a Ryd\-berg chain}. \textbf{a}: Dependence of the extended hopping parameters $\bar{J}'$ and $\bar{J}$ with the ratio $b/a$ (all parameters are normalized to the maximal value $\bar{J}_{\text{max}} \equiv |\text{max} \{\bar{J}(t)\}|$ during the cycle). The gray shaded area represents $\delta\bar{J}<0$, where the winding number of the extended SSH model is $\nu=1$ (TP: topological phase). \textbf{b}: Variation of the intracell distance $b(t)$ during the pumping cycle. Importantly, twice during the cycle the extended hopping parameters coincide, such that $\delta \bar{J}=0$. \textbf{c}: We vary the energy offset as $\Delta(t) = \Delta_\mathrm{max} \sin(2\pi t/T+\pi/2)$ such that the path in parameter space encircles the topological degeneracy point ensuring the displacement of the center of mass after a completed cycle. \textbf{d}: The left column shows the time-integrated Berry curvature of the lower band ${\cal W}_k^-$ and quasi-momentum distribution density $|f(k)|^2$ (in arbitrary units). The right column shows the spatial probability density of the wave packet in real space (on the $B$ sublattice), initially (dashed line) and after its numerical evolution for ten pumping cycles (solid line). The width of the Gaussian wave packet in momentum space, $w_k$, is doubled from the upper ($w_ka = 2\pi/100$) to the lower row ($w_k a = 2\pi/50$), leaving the displacement of the photon's center of mass $\Delta x=10\cdot0.966a$ unchanged, but leading to an increase of the dispersion, such that the fidelity per cycle  ${\cal F}(T)=0.999$ and after ten cycles ${\cal F}(10T)=0.988$ decreases to 0.997 and 0.964, respectively. Here, $T=168/\bar{J}_\text{max}$.}
    \label{GeometricalPumpingRyd}
\end{figure*}
%$\bar{J}_{\text{max}} = 5.5$MHz.

The first system we consider is an ensemble of Ryd\-berg atoms trapped in two lattices with lattice constant $a$ and offset by a distance $b$ with respect to each other as shown in Fig. \ref{fig:sketches}b~\cite{Deleseleuc2019,Lu2024}. Atoms excited to a Ryd\-berg state are known for possessing exaggerated properties, such as long lifetimes and, most importantly, extremely large transition dipole moments between neighboring Ryd\-berg states which can lead to strong long-range dipole-dipole interactions \cite{Adams2020,Browaeys2020}. The dynamics of the system is determined by the Hamiltonian \eqref{eq:Hamiltonian_QO}, where both ground and excited states of the two-level system are Ryd\-berg states. The dipole-dipole interaction is here given generally by the expression
\begin{equation}
V_{ij}=\frac{d^2}{4\pi\epsilon_0\hbar} \frac{3\cos^2{\theta_{ij}}-1}{r_{ij}^3},
\end{equation}
where $\mathbf{r}_{ij}=r_{ij}\hat{\mathbf{r}}_{ij}$ is the distance between the atoms, $\mathbf{d}=d\hat{\mathbf{d}}$ is the transition dipole moment, and $\cos{\theta_{ij}}=\hat{\mathbf{r}}_{ij}\cdot\hat{\mathbf{d}}$. To avoid breaking sublattice symmetry, we set the intra-sublattice parameters $J_{2p}$ to zero by aligning the dipoles such that they form an angle $\theta_m = \arccos(1/\sqrt{3})$ with the unit separation vector between atoms of the same sublattice (see Fig. \ref{fig:sketches}b) \cite{Deleseleuc2019}. Moreover, the dipole-dipole interactions (excitation hopping parameters) between neighboring atoms (lattice sites) dominate the dynamics, such that $\bar{J}'\approx J_1'$ and $\bar{J}\approx J_1$ with
\begin{eqnarray}
    J_1'&=&\frac{d^2}{4\pi\epsilon_0\hbar} \frac{3\cos^2{\theta_{1}'}-1}{r_{1}'^3},\\
    J_1&=&\frac{d^2}{4\pi\epsilon_0\hbar} \frac{3\cos^2{\theta_{1}}-1}{r_{1}^3}.
\end{eqnarray}
Here, $r_{1}' = \sqrt{b^2+h^2}$, $r_{1} = \sqrt{(a-b)^2+h^2}$, $\cos\theta_{1}' = \left(b\cos \theta_m-h\sin \theta_m\right)/r_{1}'$ and  $\cos\theta_{1} = \left[(a-b)\cos \theta_m+h\sin \theta_m\right]/r_{1}$. Adding an on-site potential offset between the even and odd sites, the Hamiltonian is very close to the standard Rice-Mele model \eqref{RM}. Note, however, that this is only an approximation, and the actual long-ranged character of the interactions in the Ryd\-berg system is fully taken into account in the following in our numerical calculations.

\subsubsection{Photon pumping on a Ryd\-berg chain}

While the qualitative results do not depend on the specific choice of parameters, as an illustrative example we consider a similar regime to the one realized in Ref. \cite{Deleseleuc2019}, i.e., rubidium atoms in the $60$S (ground) and $60$P (excited) states. The atoms form a chain such that the unit cells are separated by $a=12\,\mu$m and the distance between the two sublattices is $h=7.4\,\mu$m (chosen here to optimize the flatness of the time-integrated Berry curvature ${\cal W}_k^-$). The winding number of the Ryd\-berg chain can then be changed from $\nu=0$ to $1$ and vice-versa by varying $b$, which in turn changes the values of $\bar{J}$ and $\bar{J}'$. Specifically, as shown in Fig. \ref{GeometricalPumpingRyd}a, in our case when $b\lesssim 0.16 a$ the system is in a topologically trivial (dimerized) phase, while for $b\gtrsim 0.16 a$ the system possesses topologically non-trivial properties such as the existence of edge states \cite{Deleseleuc2019}.

For the photon pumping to take place, we vary time dependently the relative position of the two lattices $b(t)$ (effectively sliding sublattice $B$ keeping it parallel to sublattice $A$) and the energy offset $\Delta(t)$ as shown in Fig. \ref{GeometricalPumpingRyd}b and c, respectively. Initially, the two sublattices are arranged such that $\delta \bar{J}(t=0)<0$ and $\Delta(t=0) =\Delta_\mathrm{max}> 0$ and the photon is localized completely on sublattice $B$. Adiabatically varying $\Delta$ until it reaches the value $\Delta(t=T/2)=-\Delta_\mathrm{max}$ going past $\Delta(t=T/4)=0$, where the Hamiltonian has winding number $\nu=1$, will localize the photon on sublattice $A$. In order to surround the topological singularity we decrease $b(t)/a$ and $|\Delta|$ until we reach the point $\left(\Delta(t=3T/4)=0,\delta\bar{J}(t=3T/4)>0\right)$, where the Hamiltonian has winding number $\nu=0$, before closing the loop.

\begin{figure*}[t]
    \centering
    \includegraphics[width=\linewidth]{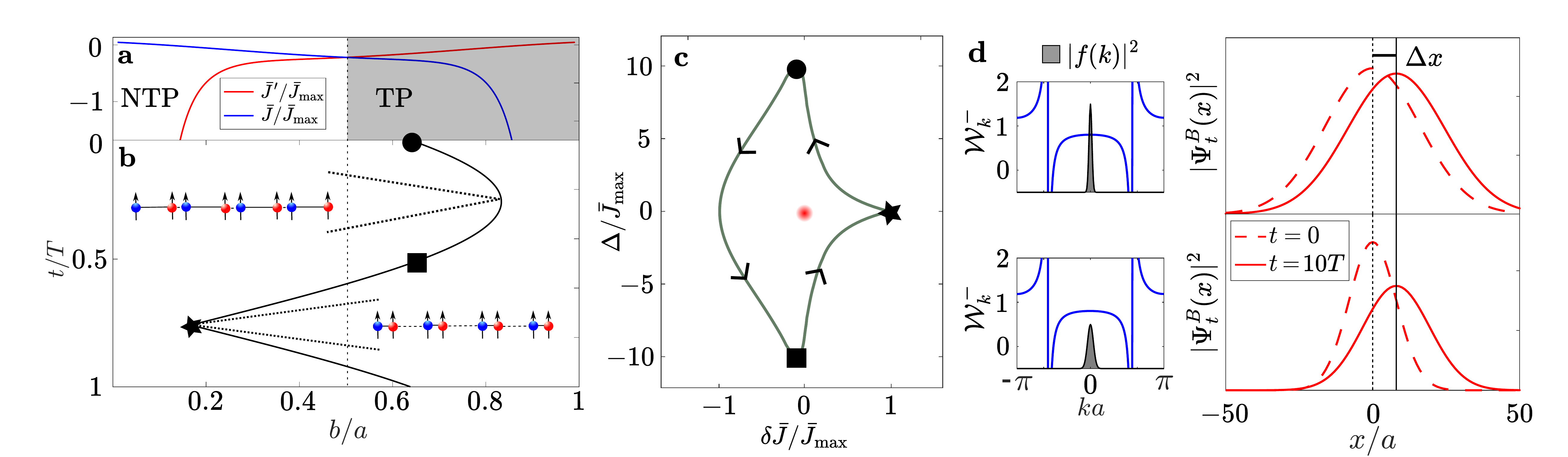}
    \caption{\textit{Photon pumping on a two-level atom chain}. \textbf{a}: Dependence of the extended hopping parameters $\bar{J}'$ and $\bar{J}$ with the ratio $b/a$ (normalized to $\bar{J}_{\text{max}} \equiv |\text{max} \{\bar{J}(t)\}|$ during the cycle). The gray shaded area represents $\delta\bar{J}<0$, where the winding number of the extended SSH model is $\nu=1$ (TP: topological phase). \textbf{b}: Variation of the intracell distance $b(t)$ during the pumping cycle. \textbf{c}: We vary the energy offset as $\Delta(t) = \Delta_\mathrm{max} \sin(2\pi t/T+\pi/2)$ such that the path in parameter space encircles the topological degeneracy point ensuring the displacement of the center of mass after a completed cycle. \textbf{d}: The left column shows the time-integrated Berry curvature of the lower band ${\cal W}_k^-$ and quasi-momentum distribution density $|f(k)|^2$ (in arbitrary units). The right column shows the spatial probability density of the wave packet in real space (on the $B$ sublattice), initially (dashed line) and after its numerical evolution for ten pumping cycles (solid line). The width of the Gaussian wave packet in momentum space $w_k$ (the same as in Fig. \ref{GeometricalPumpingRyd}) is doubled from the upper to the lower row, leaving the displacement of the photon's center of mass $\Delta x=10\cdot0.803a$ unchanged, but leading to an increase of the dispersion, such that the fidelity per cycle  ${\cal F}(T)=0.999$ and after ten cycles ${\cal F}(10T)=0.986$ decreases to 0.986 and 0.890, respectively. Here, $T=190/\bar{J}_\text{max}$.} \label{GeometricalPumpingVac}
\end{figure*}
%$\bar{J}_{\text{max}} = 1.625/\gamma$.

%We simulate numerically the dynamics by solving the Schrödinger equation with the time-dependent Hamiltonian $\hat{H}(t)$ \eqref{ExtendedRM}. We choose an initial 
We simulate the dynamics starting from a wave packet on the lower band, with a Gaussian quasi-momentum distribution $|f(k)|^2$ of width $w_k$. Here, we are able to find an optimized path in parameter space that leads to a particularly flat time-integrated Berry curvature across all the FBZ. As one can also see in Figure \ref{GeometricalPumpingRyd}d, the cyclic pumping indeed leaves the shape of the photon mostly unchanged except for its average position, which is displaced by an amount $\Delta x$. This displacement is in agreement with the one based on the time-integrated Berry curvature and the quasi-momentum distribution predicted by Equation \eqref{eq_generalized_Berry_curv}. Notably, here we are able to achieve a fidelity per cycle well above $99.9\%$, which goes down as we increase the width of the wave packet in $k$-space and dispersion plays a more prominent role.

\subsection{Long-range hopping: Atoms in free space}\label{sec:vac}

We now move to study a chain of atoms in low-lying electronic states coupled to the radiation field, such as the D-lines in alkaline atoms. Note that the main difference between this system and the Ryd\-berg one lies on the scales: we consider here an optical dipole transition instead of a microwave one (hundreds of nm versus a few cm in wavelength). The atoms form again two chains (sublattices), but in this case they are placed on a single line as shown in Fig.~\ref{fig:sketches}c. The nearest neighbor distances are now smaller, but on the same order of magnitude as the transition wavelength $\lambda_0$ \cite{Jennewein2018,Glicenstein2021}. As a consequence, as sketched in Fig. \ref{fig:sketches}c, the dipole-dipole interaction is given generally by the expression
\begin{eqnarray}\nonumber
    V_{ij} \!\!&=&\!\! \frac{3\gamma}{4}\bigg\{\!\!\left[3(\hat{\mathbf{r}}_{ij}\cdot\hat{\mathbf{d}})^2\!-\!1\right]\!\bigg[\frac{\cos k_0r_{ij}}{(k_0r_{ij})^3}+\frac{\sin k_0r_{ij}}{(k_0r_{ij})^2}\bigg] \\ &&+\left[1-(\hat{\mathbf{r}}_{ij}\cdot\hat{\mathbf{d}})^2\right]\frac{\cos k_0r_{ij}}{(k_0r_{ij})}\bigg\}, 
    \label{eq:Vijfree}
\end{eqnarray}
where $\gamma = \frac{d^2k_0^3}{3\pi\epsilon_0\hbar}$ is the single atom spontaneous decay rate, and $k_0 = \omega_0/c=2\pi/\lambda_0$ is the wave number of the transition, respectively \cite{Lehmberg1970}. This interaction can no longer be realistically approximated to be nearest neighbor only, as it was the case in the Ryd\-berg system. Thus, the corresponding extended Rice-Mele model \eqref{ExtendedRM} contains all hopping parameters $J'_{2p-1}$, $J_{2p-1}$ and $J_{2p}$. While the sublattice symmetry is in general broken in this system, we minimize this effect by choosing the distance $a$ between neighboring atoms of the same sublattice at a value $a(\theta_d)$, where $\theta_d$ is the angle between the atomic dipoles and the chain, i.e. $\cos{\theta_d}=\hat{\mathbf{r}}_{ij}\cdot\hat{\mathbf{d}}$, such that $J_{2}=0$ (see Fig. \ref{fig:sketches}c for the $\theta_d=\pi/2$ case). Given that the hopping parameters slowly decrease with the distance between atoms, the values of the subsequent symmetry-breaking hopping parameters $J_4$, $J_6$, are almost negligible, and the sublattice symmetry is only slightly broken.

\subsubsection{Photon pumping on a two-level atom chain} 
Again for illustration, here we choose a lattice constant $a=0.7 \lambda_0$ and the dipole moments oriented perpendicularly to the chain with $\theta_d=\pi/2$, which leads to a negligible intra sub-lattice hopping $J_2=0$. As was shown in \cite{Mcdonnell2022}, the winding number of the extended SSH model which is reproduced by the system when the on-site potential $\Delta=0$, again changes with the ratio $b/a$ (see Fig. \ref{GeometricalPumpingVac}a). Hence, the photon pumping is realized again by sliding the atoms trapped in the sublattice $B$ with respect to sublattice $A$, i.e., making $b(t)$ time-dependent (see Fig. \ref{GeometricalPumpingVac}b).

\begin{figure*}[t]
    \centering
    \includegraphics[width=\linewidth]{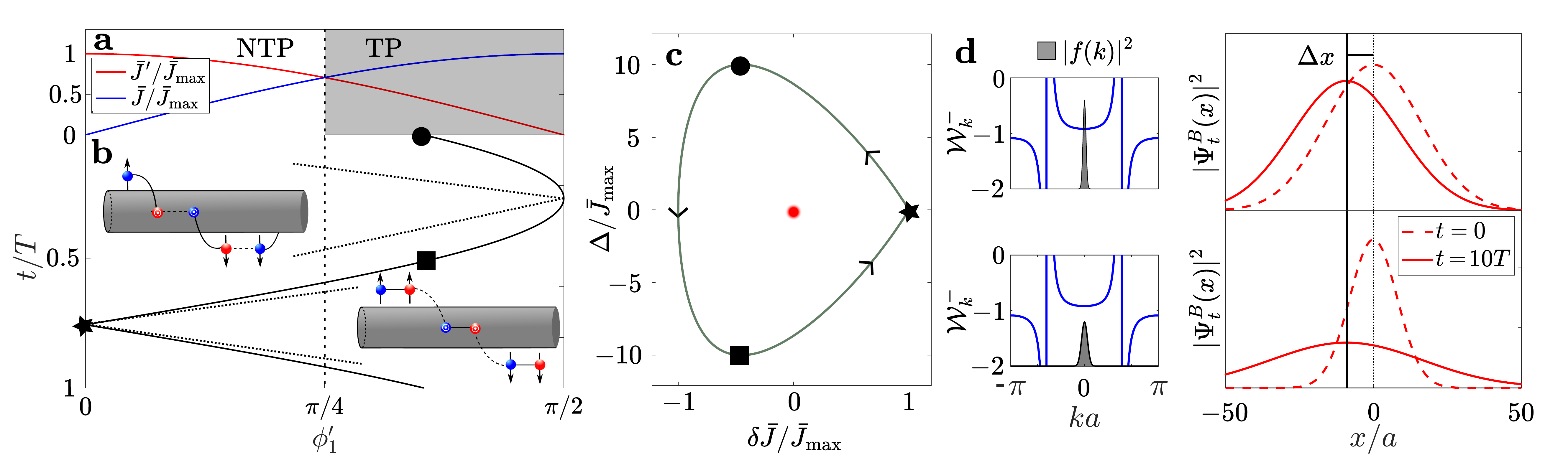}
    \caption{\textit{Photon pumping on a waveguide system}. \textbf{a}: Dependence of the extended hopping parameters $\bar{J}'$ and $\bar{J}$ with the angle $\phi'_1$ (all parameters are normalized to the maximal value $\bar{J}_{\text{max}} \equiv |\text{max} \{\bar{J}(t)\}|$ during the cycle). The gray shaded area represents $\delta\bar{J}<0$, where the winding number of the extended SSH model is $\nu=1$ (TP: topological phase). \textbf{b}: Variation of the angle $\phi'_1(t)$ during the pumping cycle. Importantly, twice during the cycle the extended hopping parameters coincide, such that $\delta \bar{J}=0$. \textbf{c}: We vary the energy offset as $\Delta(t) = \Delta_\mathrm{max} \sin(2\pi t/T+\pi/2)$ such that the path in parameter space encircles the topological degeneracy point ensuring the displacement of the center of mass after a completed cycle. \textbf{d}: The left column shows the time-integrated Berry curvature of the lower band ${\cal W}_k^-$ and quasi-momentum distribution density $|f(k)|^2$ (in arbitrary units). The right column shows the spatial probability density of the wave packet in real space (on the $B$ sublattice), initially (dashed line) and after its numerical evolution for ten pumping cycles (solid line). The width of the Gaussian wave packet in momentum space $w_k$ (the same as in Figs. \ref{GeometricalPumpingRyd} and \ref{GeometricalPumpingVac}) is doubled from the upper to the lower row, leaving the displacement of the photon's center of mass $\Delta x=10\cdot 0.919a$ unchanged, but leading to an increase of the dispersion, such that the fidelity per cycle  ${\cal F}(T)=0.998$ and after ten cycles ${\cal F}(10T)=0.962$ decreases to 0.987 and 0.540, respectively. Here, $T=150/\bar{J}_\text{max}$.}
    \label{GeometricalPumpingWG}
\end{figure*}

The pumping cycle shown in Fig. \ref{GeometricalPumpingVac}c is very similar to the one used for the Ryd\-berg chain, designed to encircle the topological degeneracy point and to maximize the region in $k$ space where ${\cal W}_k^-$ can be considered constant. As shown in Fig.~\ref{GeometricalPumpingVac}d, the long-ranged character of the couplings introduces two discontinuities in ${\cal W}_k^-$ at the so-called light lines \cite{Asenjo2017,Cech2023} which are located at $k_d a=\pm 2\pi\left(1- \frac{a}{\lambda_0}\right)$. This sets a hard constraint for the width of $|f(k)|^2$ that allows for a dispersionless topological pumping, namely $w_k\ll2k_d$. As shown Fig. \ref{GeometricalPumpingVac}d, however, dispersionless pumping can be still realized, with a fidelity per cycle as high as 99.9$\%$.

\subsection{All-to-all hopping: Waveguide system}

Finally, we discuss a chain of emitters coupled to a waveguide. This setup can be realized, for instance, with cold atoms trapped in the vicinity of a tapered optical fiber~\cite{vetsch_optical_2010,goban_demonstration_2012,nieddu_optical_2016}. In this system, the coherent interactions between atoms are mediated by the fundamental guided modes of the waveguide. Note, that we limit the discussion in this paper to cases where the coupling efficiency of the emitters to the guided modes is almost $100\%$, as it can be achieved, for example, with artificial atoms in semiconductors (quantum dots)~\cite{scarpelli_99_2019,sheremet_waveguide_2023}. In general, however, one needs to consider the coupling through the radiation or unguided modes. In the case of atoms placed around a cylindrical nanofiber (see Fig. \ref{fig:sketches}d), the interactions between atoms separated by an angle $\phi_{ij} = \phi_j-\phi_i$ and a distance $z_{ij} = z_j-z_i$ are given by
\begin{equation}
        \begin{split}
        %\Gamma_{ij}^{gd} =& \frac{\omega_a\beta_a'}{2\hbar\epsilon_0}\sum_{fl} \mathbf{d}^*_i \cdot \mathbf{e}^{(\beta_a l f)}(r_i)\mathbf{d}_j \cdot \mathbf{e}^{(\beta_a l f)*}(r_j)\\ &\times e^{il\phi_{ij}}e^{if\beta_a z_{ij}}, \\
        V_{ij} = &i\frac{\omega_0\beta'}{4\hbar\epsilon_0}\sum_{fl}\mathbf{d}_i^*\cdot \mathbf{e}^{(\beta l f)}(r_i)\mathbf{d}_j\cdot \mathbf{e}^{(\beta l f)*}(r_j)\\ &\times \text{sgn}(fz_{ij})e^{il\phi_{ij}}e^{if\beta z_{ij}},  
    \end{split}
\end{equation}
where $\mathbf{e}^{(\beta l f)}(r) = (e_r,-l e_\phi,f e_z)$ is the profile function of the fundamental guided mode with propagation constant $\beta$, polarization $l$ and propagation direction $f$ at a distance $r$ from the waveguide core, $\mathbf{d}_i = (d^r_{i},d^\phi_{i},d^z_{i})$ is the transition dipole moment of atom $i$ and $\beta' = \frac{d}{d\omega}\beta$. For simplicity, we consider the dipole moments of all emitters to point in the radial direction, and all of the atoms at the same radial distance from the waveguide core. In this case, the dipole-dipole interaction is given by the simple expression
\begin{equation}
    \begin{split}
        V_{ij} &= \frac{\gamma}{2}\cos(\phi_{ij})\sin(\beta z_{ij}),
    \end{split}
    \label{VgdGammagd}
\end{equation}
where $\gamma =\frac{2|d|^2\omega_0\beta'}{\hbar\epsilon_0} |e_r|^2$ is the rate of a photon to be emitted into the waveguide for a single atom. Moreover, we consider in particular the situation where the atomic cells are positioned as a helix around the waveguide as shown in Fig.~\ref{fig:sketches}d ~\cite{reitz_nanofiber-based_2012}. The hopping parameters in this case are given by
\begin{equation}
    \begin{split}
        &\!\!\!\!J_{2p} \!= \frac{\gamma}{2}\sin(\beta a p)\cos(p\phi_1), \\
        &\!\!\!\!J_{2p-1} \!=\! \frac{\gamma}{2}\sin\left[\beta (pa-b)\right]\cos\left(p\phi_1-\phi_1'\right), \\
        &\!\!\!\!J'_{2p-1} \!\!=\! \frac{\gamma}{2}\!\sin\!\left[\beta((p\!-\!1)a \!+\!b)\!\right]\!\cos\!\left[\!(p\!-\!1)\phi_1\!+\!\phi_1'\!\right]\! %\\
        %\Delta(t) &= \Delta_0 \sin(2\pi t/T).
    \end{split}
\end{equation}
where $\phi_1$ and $\phi_1'$ are the angle between atoms in neighboring cells of the same sublattice, and between atoms of different sublattices within the same cell, respectively. In order to ensure that the chiral symmetry is not broken, we fix the value of $a$ to $a=\pi/\beta$. We also fix the values of $b=a/2=\pi/(2\beta)$ and $\phi_1=\pi/2$, such that $J_{2p}=0$ and the remaining hopping parameters read
\begin{equation}
    \begin{split}
        J_{2p-1} &= -\frac{\gamma}{2}(-1)^{p}\cos\left(\phi_1'-\frac{p\pi}{2}\right), \\
        J'_{2p-1} &= -\frac{\gamma}{2}(-1)^{p}\sin\left(\phi_1'+\frac{p\pi}{2}\right), %\\
        %\Delta(t) &= \Delta_0 \sin(2\pi t/T).
    \end{split}
\end{equation}
having lost all dependence on the distance between the atoms in the longitudinal direction, thus being exclusively dependent on the angular offset between sublattices $\phi'_1$.

\subsubsection{Photon pumping on a waveguide system}
As shown in Fig. \ref{GeometricalPumpingWG}a, the transition from a topological to a non-topological phase occurs in this system when the angle $\phi_1'=\pi/4$, since at this point $J_{2p-1}=J'_{2p-1}$ for all values of $p$. A pumping scheme can then be established by adiabatically rotating sublattice $B$ by an angle $\phi_1'(t)$ which after a time $T$ returns to the initial configuration (Figs. \ref{GeometricalPumpingWG}b and c). As it was discussed for atoms in free space (sec Sec.~\ref{sec:vac}), due to the long-ranged character of the couplings the time-integrated Berry curvature ${\cal W}_k^-$ shows a discontinuity at the light lines at $k_da=\pm\pi/2$ (see Fig. \ref{GeometricalPumpingWG}d). This again constrains the width of the Gaussian wave packet in momentum space, which must satisfy $w_k\ll \pi/a$ in order to achieve dispersionless pumping. Here, the role of the flatness of ${\cal W}_k^-$ is made particularly clear: since the light lines are closer to $k=0$ than in the case of low-lying two-level systems, the curvature of ${\cal W}_k^-$ within the region of the Gaussian wave packet is larger, which in turn reduces the fidelity of the pumping process. Finally, we note that in the waveguide case the pumping occurs in the opposite direction compared with the two previous cases studied. This is due to the sign of ${\cal W}_k^-$ which is negative around $k=0$, while the Chern number of the band, i.e., the integral of ${\cal W}_k^-$ over the full FBZ, is the same as in all systems, namely $C_-=+1$. Note, that in principle the direction of pumping is arbitrary and depends on the orientation of the path around the singularity, i.e. the excitation moves in opposite direction by reversing the path shown in Fig.~\ref{GeometricalPumpingRyd}-\ref{GeometricalPumpingWG}c.

\begin{figure}[]
    \centering
    \includegraphics[width=\linewidth]{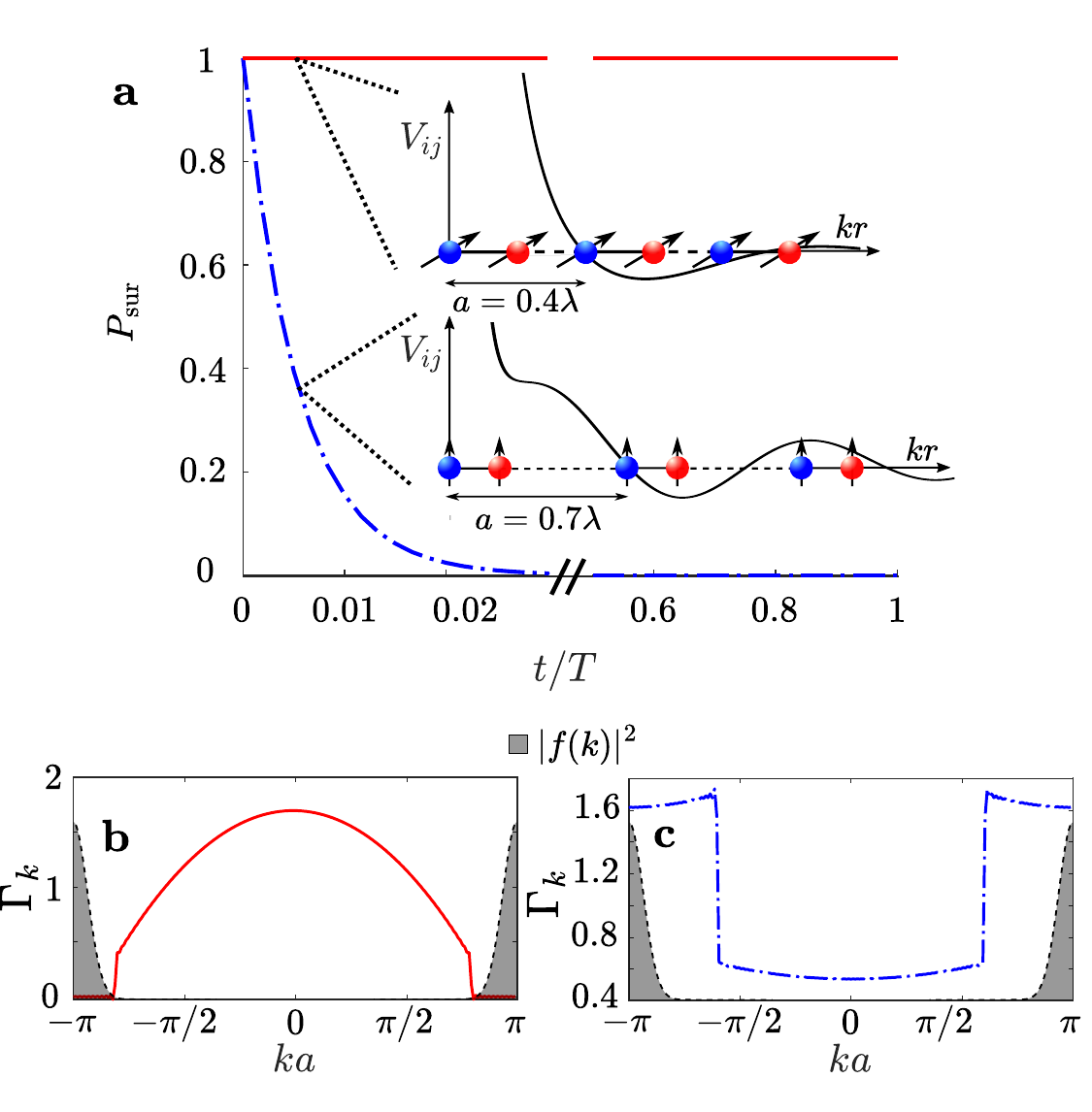}
    \caption{\textit{Collective dissipation.} \textbf{a}: Survival probability for the photon as a function of time during the pumping cycle for a chain with $a/\lambda_0=0.4$ (red solid line) and $a/\lambda_0=0.7$ (blue dash-dotted line). \textbf{b}: For $a/\lambda_0=0.4$, $|f(k)|^2$ is nonzero only in a region of the FBZ where $\Gamma_k=0$, leading to a survival probability $P_\mathrm{sur}(t)=1$ for the full cycle. \textbf{c}: The same wave packet for $a/\lambda_0=0.7$ has a large decay rate, apparent from the quick decay of the survival probability. Note, that in both cases we choose an angle $\theta_d$ such that, as discussed previously, $J_2=0$ and the sublattice symmetry is approximately conserved.}
    \label{subradianceVac}
\end{figure}

\section{Robustness against dissipation and disorder}\label{sec:robustness}

Up until now, we have shown that the topological pumping of a photon in a system with long-range couplings is possible. However, these quantum optical systems also suffer from dissipation, e.g. radiative decay of the photon into the environment, as introduced in equation \eqref{eq:MEQ}. Moreover, experimental systems have a degree of disorder due to, for example, imperfect trapping of the atoms, which leads to disorder in the hopping parameters. In this section, we will analyze the effect of these two potentially detrimental sources on the topological photon pumping protocols.

\subsection{Dissipation}\label{sec:dissipation}

Dissipation has a different character for the three platforms we have studied here. In all three, however, one can encode it in a general form as the action of the dissipator
\begin{equation}
{\cal D}(\hat{\rho})=\sum_{k=1}^N\Gamma_{k}\left[\hat{J}_k\hat{\rho} \hat{J}_k^\dag-\frac{1}{2}\left\{\hat{J}_k^\dag \hat{J}_k,\hat{\rho}\right\}\right],
\label{Dissipator}
\end{equation}
where $\hat{J}_k$ is a so called jump operator and $\Gamma_k$ is the rate at which this jump occurs. Here, the \textit{jumps} are emissions of a photon into the environment.

For the chain of Ryd\-berg atoms, the main emission channel is the independent decay with the same rate , i.e. $\Gamma_k=\Gamma$, from the Ryd\-berg states into the atomic ground state \cite{Sibalic2017}. As an illustration, the comparison between the lifetime of the 60P rubidium Ryd\-berg state, $\tau_R\approx500\,\mu$s, and an optimal pumping period of $T \approx 22\,\mu$s means that the pumping process is likely to survive dissipation at least for a few cycles. However, to overcome this constraint one can think of, for example, leveraging the remarkable lifetimes afforded by circular Ryd\-berg states \cite{Hoelzl2024}, which exhibit exceptional decoupling from optical dipole transitions.

In the second case we considered, i.e., atoms in low-lying energy states, the nearest neighbor distances are comparable to $\lambda_0$ and the dissipation acquires a collective character. Here, the dissipator reads
\begin{equation}
{\cal D}(\hat{\rho})=\sum_{i,j=1}^N\Gamma_{ij}\left[\hat{\sigma}_i\hat{\rho} \hat{\sigma}_j^\dag-\frac{1}{2}\left\{\hat{\sigma}_i^\dag \hat{\sigma}_j,\hat{\rho}\right\}\right],
\label{Dissipator_coll}
\end{equation}
where the $\Gamma_{ij}$ is, in general, not diagonal and reads
\begin{eqnarray}\nonumber
    \Gamma_{ij} \!\!&=&\!\! \frac{3\gamma}{2}\bigg\{\!\left[3(\hat{\mathbf{r}}_{ij}\cdot\hat{\mathbf{d}})^2\!-\!1\right]\!\bigg[\frac{\sin k_0r_{ij}}{(k_0r_{ij})^3}\!-\!\frac{\cos k_0r_{ij}}{(k_0r_{ij})^2}\bigg] \\ &&+\left[1-(\hat{\mathbf{r}}_{ij}\cdot\hat{\mathbf{d}})^2\right]\frac{\sin k_0r_{ij}}{(k_0r_{ij})}\bigg\},
\label{eq:Gammaijfree}
\end{eqnarray}
i.e., as in the case of the dipole-dipole interactions \eqref{eq:Vijfree}, it depends strongly on the ratio between the interatomic distance and the transition wavelength, the precise external geometry of the system and the orientation of the transition dipole moments \cite{Lehmberg1970}. The (collective) jump operators and rates can be found by diagonalizing this matrix, such that $\Gamma_k=\sum_{ij}M_{ki}\Gamma_{ij}M^*_{jk}$ and $\hat{J}_k=\sum_j M_{kj}\hat{\sigma}_j$. Here one can see that the jump operators are then in general superpositions of all single atom spin operators $\hat{\sigma}_j=\ket{g}_j\!\!\bra{e}$, with rates that can be larger (superradiant) or smaller (subradiant) than the single atom decay rate. In the regime we consider here, a small number of superradiant states with $N\gamma>\Gamma_k$ are present. The region of the FBZ where these superradiant states appear, however, depends strongly on the specific value of the ratio between nearest neighbor distance and $\lambda_0$. In particular, due to the large sublattice energy shifts $\Delta$ that are present for most of the pumping cycle, the two sublattices are effectively decoupled dynamically, which means that the reduced distance that determines the dissipation properties is $a/\lambda_0$. As can be observed in Figs. \ref{subradianceVac}b and c, the superradiant rates can be found at the center and the border of the FBZ for $a/\lambda_0<1/2$ and $a/\lambda_0>1/2$, respectively. This in turn means that, as shown in Fig. \ref{subradianceVac}b, choosing a value $a/\lambda_0<1/2$, one can create a wave packet centered at $k=\pm\pi/a$ with support exclusively in the subradiant region, which in turn leads to a survival probability of the photon $P_\mathrm{sur}$ equal to one for all experimentally relevant times. For comparison, the same wave packet created on a chain where $a/\lambda_0>1/2$ will decay fast (in the example we show in Fig. \ref{subradianceVac}a with $a/\lambda_0=0.7$, $\Gamma_\mathrm{eff}=1.6183\gamma$).

Finally, in the case of an ensemble of atoms coupled to a waveguide, the dissipation is described by the same dissipator \eqref{Dissipator_coll}, where now the dissipation coefficient matrix is given by
\begin{equation}
        \begin{split}
        \Gamma_{ij} =& \frac{\omega_0\beta'}{2\hbar\epsilon_0}\sum_{fl} \mathbf{d}^*_i \cdot \mathbf{e}^{(\beta l f)}(r_i)\mathbf{d}_j \cdot \mathbf{e}^{(\beta l f)*}(r_j)\\ &\times e^{il\phi_{ij}}e^{if\beta z_{ij}}.
    \end{split}
\end{equation}
In the parameter regime we have considered in the previous section, this expression reduces to
\begin{equation}
    \begin{split}
        %V_{ij} &= \frac{\gamma}{2}\cos(\phi_{ij})\sin(\beta z_{ij}),% \\
        \Gamma_{ij} &= \gamma \cos(\phi_{ij})\cos(\beta z_{ij}).
    \end{split}
    \label{Gammagd}
\end{equation}
The diagonalization of this matrix shows that only two collective decay channels possess non-zero decay rates. Since the Gaussian wave packet that constitutes our initial state has no overlap with these two superradiant states, dissipation in this case has virtually no effect on the topological pumping process. Note, that again we have assumed that the dissipation only occurs into the guided modes on the waveguide, neglecting the effect of unguided modes.

\begin{figure}[]
    \centering
    \includegraphics[width=\linewidth]{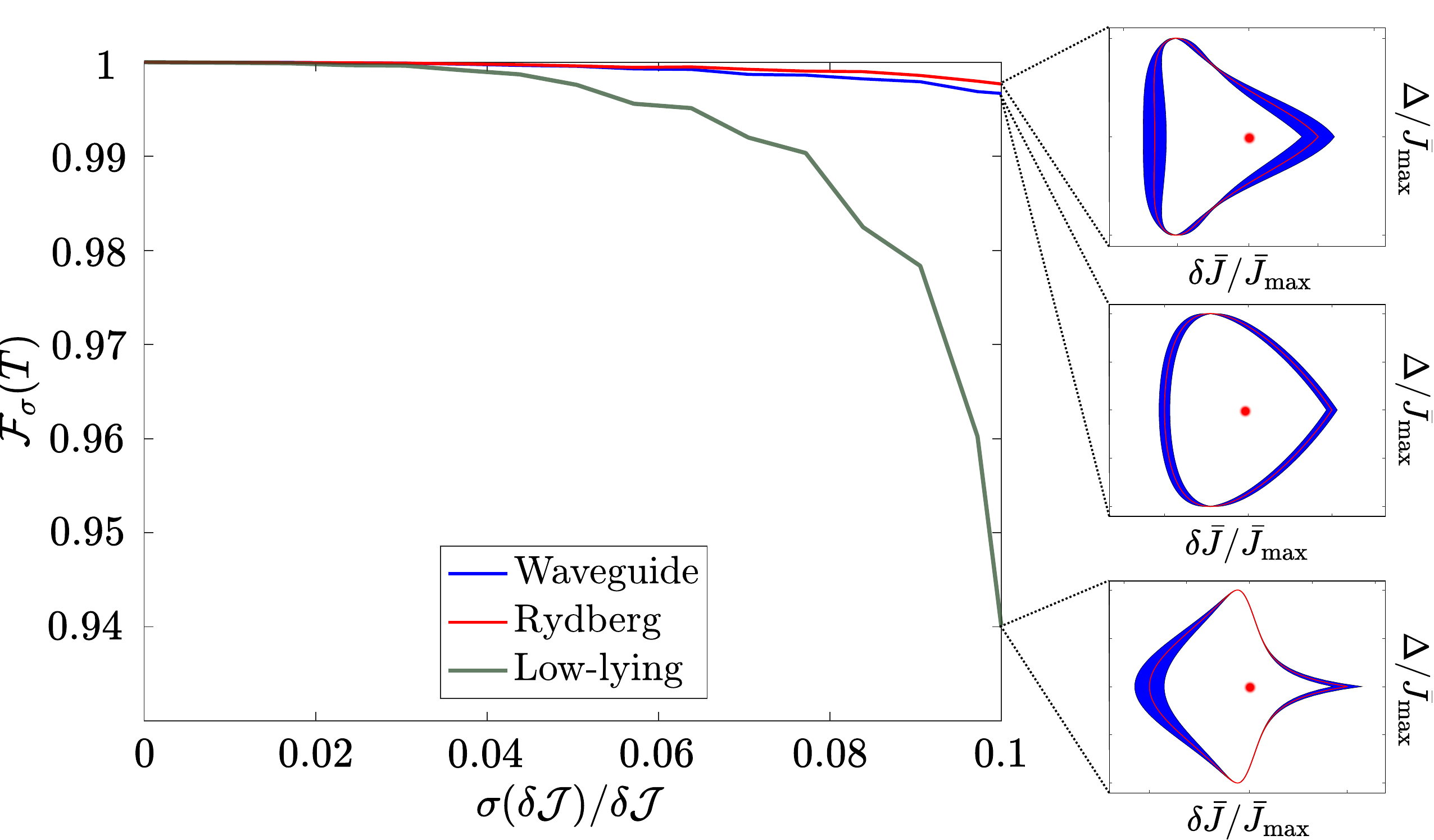}\caption{\textit{Robustness against disorder}. Fidelity after one pumping cycle $\mathcal{F}_\sigma(T)$ between the disordered and non-disordered states as a function of the time-integrated disorder in the extended hopping parameters $\sigma(\delta \mathcal{J})/\delta{\mathcal J}$. The insets show the disordered paths in parameter space for the three systems with a time-integrated disorder of 10\%. The unperturbed path in parameter space is shown by the red line, while the blue shaded area shows the propagated disorder given a Gaussian disorder in each emitter position in the chain.} 
    \label{Fidelity}
\end{figure}

\subsection{Disorder}\label{sec:disorder}

Finally, we analyze the robustness against disorder of the topological pumping realized on the three quantum optical platforms considered. In particular, we consider the unavoidable disorder on the positions of the atoms by adding random Gaussian noise to each atom position. Contingent upon the dependence of the dipole-dipole interactions, this leads to a degree of disorder on the path in parameter space given by the curve $\xi = (\delta \bar{J},\Delta)$. How the disorder in the individual emitter positions translates into disorder in the path in parameter space is explained in detail in Appendix \ref{app:disorder}. Considering that the disorder varies along the path, we introduce the time-integrated function
\begin{equation}
    \delta \mathcal{J} \equiv \int_0^T dt\, \delta \bar{J}(t),
\end{equation}
and its standard deviation
\begin{equation}
    \sigma(\delta \mathcal{J}) \equiv \int_0^T dt\, \sigma\left[\delta \bar{J}(t)\right],
\end{equation}
as a common measure of disorder for the three systems. Hence, in order to get a fair comparison, we choose the same value of the time-integrated disorder $\sigma (\delta\mathcal{J})/\delta\mathcal{J}$ for all systems. Note, however, that as a result of propagation of disorder the Ryd\-berg system with short range couplings requires a larger disorder in the position compared to the two other systems to reach a equivalent disorder in the path.

To measure the effect of disorder, we calculate the fidelity
\begin{equation*}
\mathcal{F}_\sigma(T) = \left(\mathrm{Tr}\sqrt{\sqrt{\rho_T^0}\rho_T^{\sigma}\sqrt{\rho_T^0}}\right)^2,    
\end{equation*}
where $\rho_T^{\sigma}$ is the disordered state averaged over a number of realizations of random Gaussian disorder in the emitter positions and $\rho_T^0=|\Psi_T^0\rangle\langle\Psi_T^0|$ is the non-disordered state after one cycle. Figure \ref{Fidelity} shows this fidelity for the three different systems as a function of the time-integrated disorder $\sigma (\delta\mathcal{J})/\delta\mathcal{J}$ up to 10\%. We can observe that the pumping is indeed very robust in all three cases, with the atoms in low-lying states being the most affected, likely due to the slight breaking of the sublattice symmetry present in this case, i.e. the presence of $J_{2p}\neq 0$ terms, which effectively double the amount of couplings affected by the disorder.

\section{Conclusions and outlook}
This study establishes for the first time a framework for topological photon pumping in quantum optical systems with long-range couplings. 
This framework allows for the controlled and robust transport of photonic excitations over a one-dimensional emitter chain. Notably, the atoms are only displaced over short distances within a single unit cell, while the excitation is transported with high fidelity over many lattice sites.
%This framework allows for a controlled and robust transport with high fidelity of photonic excitations over many lattice sites of a one-dimensional emitter chain. 
We discuss its realization in different experimental platforms, such as Ryd\-berg atoms, in a dense atomic lattice gas, and atoms coupled to waveguides. For each system we establish a counter-intuitive protocol that displaces only half of the atoms, but allows to transport the photonic excitation with a fidelity reaching 99.9\%. Our results underscore that this scheme not only shifts the center-of-mass of the excitation, but also maintains the shape of the wave packet (dispersionless) and we show that it is robust against both local disorder and dissipation.

While in our work we found a parameter trajectory that is suitable for realizing pumping, we did not use any optimization procedure to obtain an optimal path. Depending on the context, such strategies could be employed to optimize the displacement $\Delta x$ per cycle, or the speed  of the pumping, while maintaining adiabaticity, e.g. faster when gap is large and slower when gap is narrow. Moreover, the paramaters of the system could be optimized to realize flatter time-integrated Berry curvatures which in turn minimize the dispersion.

While our results are obtained in 1D systems, these concepts can be generalized to higher dimensional systems \cite{lohse_exploring_2018}. Finally, such charge pump could be extended to intriguing \textit{spin} pumping schemes that include different light polarizations corresponding to coupling to different transitions. This could mimic a Fu-Kane pump~\cite{fu_time_2006} which in condensed matter systems allow for a spin transport without charge transport. 

While this manuscript was under review, we became aware of an experimental realization of a charge pump with Rydberg excitations \cite{trautmann2024}.

\section{Acknowledgements}
 M.S. thanks I.B. Spielman \& H.-I Lu for their introduction to the topic of topological charge pumps. M.S. acknowledges support by Young Researcher Grant MajorSuperQ MSCA 0000048. The authors acknowledge support by the state of Baden-Württemberg through bwHPC and the German Research Foundation (DFG) through grant no INST 40/575-1 FUGG (JUSTUS 2 cluster). We acknowledge support by Open Access Publishing Fund of University of Tübingen. The research leading to these results has received funding from the Deutsche Forschungsgemeinsschaft (DFG, German Research Foundation) under Project No. 452935230 and the Research Units FOR 5413/1, Grant No. 465199066 and FOR 5522/1, Grant No. 499180199.

\appendix

\section{Generalized expression for $\Delta x$}\label{app:displacement}

We consider the state of the system throughout the dynamics to be a superposition of Bloch waves, i.e.
\begin{equation*}
\ket{\Psi_t(x)} = \frac{1}{\sqrt{N}}\sum_{k=-\pi/a}^{\pi/a} f(k)\ket{\psi_{nk}(x)},
\end{equation*}
where $N$ is the number of unit cells, $f(k)$ is the quasi-momentum distribution normalized such that
\begin{equation*}
    \int_\mathrm{FBZ}|f(k)|^2\mathrm{d}k=\frac{2\pi}{a},
\end{equation*}
and
\begin{equation*}
    \ket{\psi_{nk}(x)}=\frac{1}{\sqrt{N}}e^{ikx}\ket{x}\otimes\ket{u_{nk}(t)}.
\end{equation*}
In the limit of an infinite system, i.e., $N \to \infty$, $k$ becomes a continuous variable. Thus, we can substitute the sum over $k$, $\sum_{k = -\pi/a}^{\pi/a}$, by the corresponding integral, $\frac{Na}{2\pi} \int_\mathrm{FBZ}\mathrm{d}k$, such that the state reads
\begin{equation*}
    \ket{\Psi_t(x)} = \frac{a}{2\pi}\int_{\text{FBZ}} \mathrm{d}k f(k)e^{ikx}\ket{x}\otimes\ket{u_{nk}(t)}.
\end{equation*}
The expectation value of the position as a function of time is given by
\begin{equation*}
\begin{split}
    \langle\hat{x}\rangle_t&=\sum_{x=-\infty}^\infty\bra{\Psi_t(x)}\hat{x}\ket{\Psi_t(x)}\\&=\left(\frac{a}{2\pi}\right)^2\sum_{x}\int_{\text{FBZ}}\int_{\text{FBZ}}\mathrm{d}k \mathrm{d}k' f(k)f^*(k')\\&\times\bra{u_{nk'}(t)}xe^{ix(k-k')} \ket{u_{nk}(t)}. 
\end{split}
\end{equation*}
Using partial integration we get
\begin{equation*}
\begin{split}
    &\langle\hat{x}\rangle_t = i\left(\frac{a}{2\pi}\right)^2\sum_{x}\int_{\text{FBZ}}\mathrm{d}k'f^*(k')\bra{u_{nk'}(t)}e^{-ik'x}\\&\times\!\int_{\text{FBZ}}\!\!\!\mathrm{d}k e^{ikx}\lbrace\left[\partial_k f(k) \right] \ket{u_{nk}(t)}+f(k)\partial_k\ket{u_{nk}(t)}\rbrace.
\end{split}
\end{equation*}
Using the Fourier decomposition of the Dirac delta function $\sum_{x=-\infty}^\infty e^{i(k-k')x}=\frac{2\pi}{a}\delta(k-k')$, we obtain
\begin{equation*}
\begin{split}
    \langle\hat{x}\rangle_t &=  \frac{ia}{2\pi}\!\int_{\text{FBZ}}\mathrm{d}k\bigg[f^*(k)\partial_kf(k)\\
    &+|f(k)|^2\bra{u_{nk}(t)}\partial_ku_{nk}(t)\rangle\!\bigg]\!.
\end{split}
\end{equation*}
After one full cycle of time $T$, the center of mass of the wave packet is displaced by
\begin{equation*}
\begin{split}
    \Delta x=&\langle \hat{x}\rangle_T-\langle\hat{x}\rangle_0 =  \\=&  \frac{a}{2\pi}\int_\text{FBZ}\mathrm{d}k|f(k)|^2\left[A_{nk}(T)-A_{nk}(0)\right],    
\end{split}
\end{equation*}
where we have introduced the Berry connection $A_{nk}(t) = i\bra{u_{nk}(t)}\partial_ku_{nk}(t)\rangle$. Considering that, after a full adiabatic cycle, the time-evolved eigenstates only pick up the so-called geometric or Berry phase, i.e. $\ket{u_{nk}(T)}=e^{i\gamma(k)} \ket{u_{nk}(0)}$, one can easily find that
\begin{equation*}
\begin{split}
    \Delta x   &=  -\frac{a}{2\pi}\int_\mathrm{FBZ}  |f(k)|^2\partial_k \gamma(k) \mathrm{d}k\\&=
    \frac{a}{2\pi} \int_0^T \mathrm{d}t \int_\mathrm{FBZ}  |f(k)|^2\Omega_{tk} \mathrm{d}k,
\end{split}
\end{equation*}
as given in the main manuscript.

%\begin{equation*}
%\begin{split}
%    \Delta x=&\langle \hat{x}\rangle_T-\langle\hat{x}\rangle_0 = \int_0^T\mathrm{d}t\,\partial_t \langle \hat{x} \rangle_t \\=&  \frac{a}{2\pi}\int_0^T\mathrm{d}t\int_\text{FBZ}\mathrm{d}k|f(k)|^2\partial_tA_{nk}(t),    
%\end{split}
%\end{equation*}
%where we have introduced the Berry connection $A_{nk}(t) = i\bra{u_{nk}(t)}\partial_ku_{nk}(t)\rangle$. The integral may now be expressed in term of the Berry curvature of the band $\Omega_{tk} = \partial_tA_{nk}-\partial_kA_{nt} = -2\text{Im}\{\langle\partial_t u_{nk}(t)|\partial_k u_{nk}(t)\rangle\}$ as 
%\begin{equation*}
%\begin{split}
%    &\Delta x = \frac{a}{2\pi} \int_0^T \mathrm{d}t \int_\mathrm{FBZ}  |f(k)|^2\Omega_{tk} \mathrm{d}k \\&+ \frac{ia}{2\pi}\int_0^T \mathrm{d}t \int_\mathrm{FBZ} \mathrm{d}k|f(k)|^2\partial_k\bra{u_{nk}(t)}\partial_tu_{nk}(t)\rangle. 
%\end{split}
%\end{equation*}
%Using partial integration on the last integral over $k$ and using the periodicity of the function $\bra{u_{nk}(t)}\partial_tu_{nk}(t)\rangle$ in the FBZ, this term vanishes and we end up with the final expression for the displacement of the center of mass

\section{Propagation of disorder} \label{app:disorder}
Here we study how the disorder in the position of the emitters in the chain translates to disorder in the path in parameter space given by the curve $\xi(t) = (\delta \bar{J}(t),\Delta(t))$. Consider the position of all emitters in a 1D chain to contain some Gaussian distributed disorder with standard deviation $\sigma (\mathbf{r})$ around the equilibrium position $\mathbf{r}$. The corresponding standard deviation of the distribution of distances between any two emitters in the chain $r_{ij}=|\mathbf{r}_j-\mathbf{r}_i|$ is then given by 
\begin{equation*}
    \sigma (r_{ij}) = \sqrt{2}\frac{\mathbf{r}_{ij}\cdot\sigma(\mathbf{r})}{r_{ij}}.
\end{equation*}
Such a Gaussian disorder in the distance $r_{ij}$ translates into a disorder in the hopping parameters $J(r_{ij},t)$ as
\begin{equation*}
    \sigma\left[J(r_{ij},t)\right] = \frac{\partial J(r_{ij},t)}{\partial r_{ij}} \sigma (r_{ij}) .
\end{equation*}
Due to the form of the extended hopping parameters $\bar{J}'$ and $\bar{J}$, the disorder propagates when considering the full path in parameter space. Using the theory of error propagation we derive the disorder in the extended hopping parameter difference $\delta \bar{J}$, which is given by
\begin{equation*}
    \sigma (\delta \bar{J}) = \sqrt{\sigma(\boldsymbol{\mathfrak{J}})\cdot \sigma(\boldsymbol{\mathfrak{J}})+\frac{1}{2}\sum_{p\neq q}^{2\lfloor N/4 \rfloor} K_{pq}(t)},
\end{equation*}
where we have introduced the extended hopping vector $\boldsymbol{\mathfrak{J}}$ with odd components $\mathfrak{J}_{2p-1} = (-1)^{(p+1)}J'_{2p-1}$ and even components $\mathfrak{J}_{2p} = (-1)^pJ_{2p-1}$ and the covariance matrix $K = \sigma (\boldsymbol{\mathfrak{J}}^T)\sigma (\boldsymbol{\mathfrak{J}})$.

\bibliographystyle{quantum}

\end{document}